\documentclass[natbib]{svjour3}                     % onecolumn (standard format)
\smartqed  % flush right qed marks, e.g. at end of proof
\usepackage{graphicx}
\usepackage{amssymb,amsmath}
 \usepackage{aps-bibstyle}  % use this style if you don't use BibTeX.
\DeclareOldFontCommand{\rm}{\normalfont\rmfamily}{\mathrm}

\renewcommand{\d}{\mbox{d}}
\newcommand{\ssmall}{\mbox{\scriptsize s}}
\newcommand{\dsmall}{\mbox{\scriptsize l}}

\newcommand{\gsim}{\raise 
        -2.truept\hbox{\rlap{\hbox{$\sim$}}\raise5.truept 
        \hbox{$>$}\ }}
\newcommand{\lsim}{\raise 
        -2.truept\hbox{\rlap{\hbox{$\sim$}}\raise5.truept 
        \hbox{$<$}\ }}  
\newcommand{\minmag}{\raise-2.truept\hbox{\rlap{\hbox{$<$}}\raise 
        6.truept\hbox 
        {$>$}\ }}

\begin{document}

\title{Arc Statistics%\thanks{Grants or other notes
}
\subtitle{}


\author{Meneghetti M.         \and
        Bartelmann M. \and 
        Dahle H. \and
        Limousin M. %etc.
}


\institute{M. Meneghetti \at
              INAF - Osservatorio Astronomico di Bologna, Via Ranzani 1, 40127, Bologna, Italy \\
              Tel.: +39-051-2095815\\
              Fax: +39-051-2095700\\
              INFN - Sezione di Bologna, viale Berti Pichat 6/2, 40127,
Bologna, Italy \\
              \email{massimo.meneghetti@oabo.inaf.it}           %  \\
           \and
           M. Bartelmann \at
              U. Heidelberg, ZAH, ITA, Philosophenweg 12, 69120 Heidelberg, Germany
           \and
           H. Dahle \at
           	  Institute of Theoretical Astrophysics, University of Oslo, P.O. Box 1029, Blindern, NO-0315 Oslo, Norway
	  	   \and
		   M. Limousin \at
		   Aix Marseille Universit\'e, CNRS, LAM (Laboratoire d'Astrophysique de Marseille), UMR 7326, 13388, Marseille, France
              }

\date{Received: date / Accepted: date}

\maketitle

\begin{abstract}
The existence of an {\em arc statistics} problem was at the center of a strong debate in the last fifteen years. With the aim to clarify if  the optical depth for {\em giant} gravitational arcs by galaxy clusters in the so called concordance model is compatible with observations, several studies were carried out which helped to significantly improve our knowledge of strong lensing clusters, unveiling their extremely complex internal structure. In particular, the abundance and the frequency of strong lensing events like gravitational arcs turned out to be a potentially very powerful tool to trace the structure formation. However, given the limited size of observational and theoretical data-sets, the power of arc statistics as a cosmological tool has been only minimally exploited so far. On the other hand, the last years were characterized by significant advancements in the field, and several cluster surveys that are ongoing or planned for the near future seem to have the potential to make arc statistics a competitive cosmological probe. Additionally, recent observations of anomalously large Einstein radii and concentrations in galaxy clusters have reinvigorated the debate on the arc statistics problem. In this paper, we review the work done so far on arc statistics, focussing on what is the lesson we learned and what is likely to improve in the next years.     
\keywords{Cosmology \and Galaxy clusters \and Gravitational Lensing}
\end{abstract}

\section{Gravitational arcs in galaxy clusters}
\label{intro}
\subsection{Phenomenology and immediate conclusions}

Strong gravitational lensing in galaxy clusters was first detected by \cite{1987A&A...172L..14S} and \cite{LY89.1}. They found extended, faint, blueish arc-like images in the cores of the galaxy clusters Abell~370 ($z = 0.373$, ClG~0237.3$-$0148) and ClG~2244$-$0221 ($z = 0.328$, also known as 2E~2244.6$-$0221). Several viable explanations were proposed for these objects, among them star formation behind galactic bow shocks \citep{BE87.1}, but also gravitational lensing of background galaxies \citep{PA87.1}. The latter hypothesis was confirmed when the redshift of the arc in Abell~370 was measured and found to be $z = 0.724$, substantially higher than the cluster's redshift \citep{SO88.1}.

\begin{figure}
  \includegraphics[width=1.0\hsize]{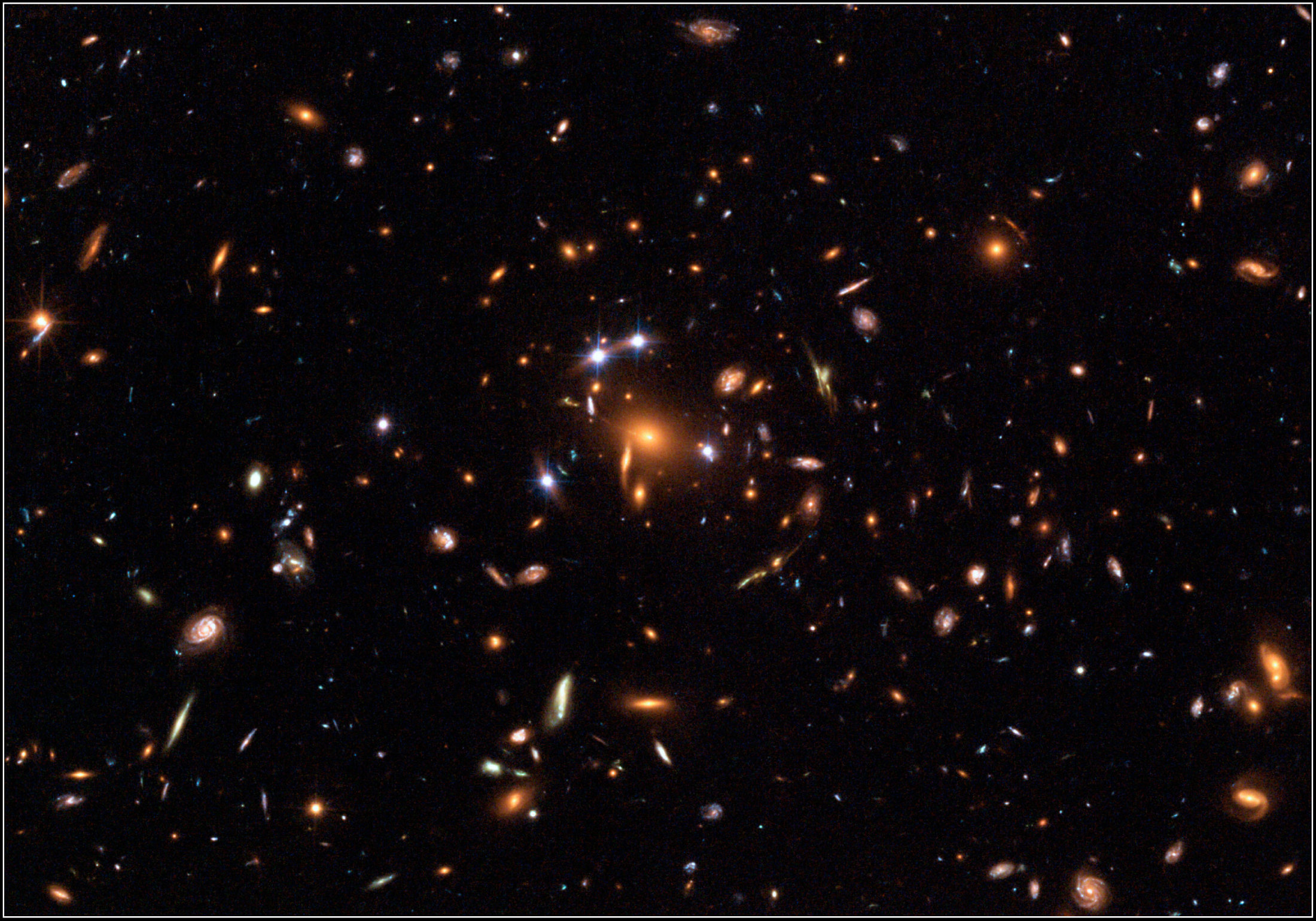}
\caption{This HST image of the cluster SDSS J~1004+4112 \citep{SH05.1} shows arcs, multiply imaged galaxies and a quadruply lensed quasar with a maximum image separation of $14.62$ arc seconds \citep{IN03.1}.}
\label{fig:11}       % Give a unique label
\end{figure}

It was quickly recognised that gravitational arcs provide important information on the core structure of galaxy clusters and their mass distribution. It was unclear at the time how the dark matter was distributed and whether the X-ray surface-brightness profiles, which typically show a flat core with a radius near $\simeq200\,h^{-1}\,\mathrm{kpc}$, were representative for the dark-matter profiles ($h$ here denotes the usual $H_0/100$, where $H_0$ is the Hubble parameter at the present epoch). Arcs were soon found to reveal the following about clusters: (1) Cluster mass distributions cannot typically be axially symmetric, because large counter-arcs would otherwise be expected \citep{GR88.1, KO89.1}. (2) The substantial amounts of dark matter in galaxy clusters cannot be attached to the galaxies because arcs would then have much smaller curvature radii \citep{HA89.2, BE90.1}. Particularly striking were the detections of ``straight arcs'' in two clusters \citep{PE91.1, MA92.1, PI96.1} because they visually demonstrated the need for substantial concentrations of dark matter with very high mass-to-light ratio \citep{KA92.1}. (3) Clusters need to have steep density profiles, because arcs would be substantially thicker otherwise \citep{HA89.1}. For clusters to be strong lenses, their central convergence $\kappa$ has to be close to unity, but for arcs to be thin, the convergence at their locations has to be around $0.5$. From cluster centres to the arc radii of typically $10''\ldots30''$, the $\kappa$ profile must thus fall by approximately a factor of two. Cluster core radii, if they exist, must thus be substantially smaller than the X-ray core radii, which was also confirmed by the detection of ``radial arcs'' \citep{FO92.1, MI93.1, ME93.1}.

\subsection{Axially-symmetric lens models}
\label{sect:axsym}

These immediate statements on gravitational arcs are quite generic and can be easily understood. For an axially-symmetric lens, the lens equation is effectively one-dimensional,
\begin{equation}
  \beta(\theta) = \theta-\alpha(\theta)\;,
\label{eq:01}
\end{equation}
where $\theta$ is the angular separation from an optical axis coinciding with the symmetry axis of the lens, and $\beta(\theta)$ is the angular separation of the (unseen) source from the optical axis whose image appears at $\theta$. An axially-symmetric lens with a line-of-sight projected surface-mass density $\Sigma(\theta)$ produces the deflection angle
\begin{equation}
  \alpha(\theta) = \frac{m(\theta)}{\theta}\;,
\label{eq:02}
\end{equation}
where $m(\theta)$ is proportional to the projected, cumulative mass profile of the lens,
\begin{equation}
  m(\theta) = 2\int_0^\theta\theta'\d\theta'\,\kappa(\theta')\;,
\label{eq:03}
\end{equation}
and the so-called convergence $\kappa(\theta)$ is the geometrically scaled surface-mass density,
\begin{equation}
  \kappa(\theta) = \frac{\Sigma(\theta)}{\Sigma_\mathrm{cr}(z_\mathrm{l}, z_\mathrm{s})}\;.
\label{eq:04}
\end{equation} 
The critical surface-mass density $\Sigma_\mathrm{cr}(z_\mathrm{l}, z_\mathrm{s})$ introduced here depends on the angular-diameter distances between the observer and the lens $D_\mathrm{l}$, the observer and the source $D_\mathrm{s}$ and between the lens and the source, $D_\mathrm{ls}$,
\begin{equation}
  \Sigma_\mathrm{cr}(z_\mathrm{l}, z_\mathrm{s}) = \frac{c^2}{4\pi G}\frac{D_\mathrm{s}}{D_\mathrm{l}D_\mathrm{ls}}\;.
\label{eq:05}
\end{equation}

Strong distortions occur where the lens mapping (\ref{eq:01}) becomes singular, i.e.~where its Jacobian determinant vanishes,
\begin{equation}
  \det\left(\frac{\partial\beta_i}{\partial\theta_j}\right) = \frac{\beta(\theta)}{\theta}\frac{\d\beta(\theta)}{\d\theta} = 0\;.
\label{eq:06}
\end{equation}
Combining (\ref{eq:01}), (\ref{eq:02}) and (\ref{eq:03}), we immediately find
\begin{equation}
  \frac{\beta(\theta)}{\theta}\frac{\d\beta(\theta)}{\d\theta} =
  \left(1-\frac{\alpha(\theta)}{\theta}\right)\left(1-\frac{\d\alpha(\theta)}{\d\theta}\right) =
  \left(1-\frac{m(\theta)}{\theta^2}\right)\left(1-\frac{\d}{\d\theta}\frac{m(\theta)}{\theta}\right)\;.
\label{eq:07}
\end{equation}
First of all, this result shows immediately that strong distortions, hence also arcs, can occur in two physically distinct locations, either where the scaled, enclosed mass $m(\theta)$ satisfies
\begin{equation}
  m(\theta) = \theta^2
\label{eq:08}
\end{equation}
or where the slope of the mass profile satisfies
\begin{equation}
  \frac{\d}{\d\theta}\frac{m(\theta)}{\theta} = 2\kappa(\theta)-\frac{m(\theta)}{\theta^2} = 1\;.
\label{eq:09}
\end{equation}
Both conditions define the so-called critical curves which, in the idealised case of axially-symmetric lenses, form circles around the lens centre that may degenerate to points. They have an intuitive meaning that is most useful for the interpretation of strong gravitational lensing. Equation (\ref{eq:08}) states that one type of strongly distorted image constrains the total scaled mass interior to it, while (\ref{eq:09}) shows that another type of image occurs where the scaled mass profile has an appropriate slope.

Images occurring at the locations defined by (\ref{eq:08}) and (\ref{eq:09}) can easily be distinguished. Closer inspection of the lens mapping shows that images satisfying (\ref{eq:08}) are tangentially distorted, while the condition (\ref{eq:09}) produces radially distorted images reminding of spokes pointing away from the cluster centre. Tangential arcs thus allow estimates of the total, scaled, enclosed lens mass, while radial arcs constrain the slope of the mass profile. We thus expect a typical strong gravitational lens to produce tangential and radial critical curves with angular radii $\theta_\mathrm{t}$ and $\theta_\mathrm{r}$, respectively.

The inverse of the Jacobian determinant at an image location is the magnification $\mu$ of the image. For tangential arcs, the first factor on the right-hand side of (\ref{eq:07}) vanishes. The magnification of tangential arcs in the radial direction is thus given by the inverse second factor,
\begin{equation}
  \mu_\mathrm{r} =
  \left.\left(1-\frac{\d}{\d\theta}\frac{m(\theta)}{\theta}\right)^{-1}\right|_{\theta=\theta_\mathrm{t}} =
  \left.\left(1-2\kappa+\frac{m(\theta)}{\theta^2}\right)^{-1}\right|_{\theta=\theta_\mathrm{t}} =
  2\left[1-\kappa(\theta_\mathrm{t})\right]^{-1}\;,
\label{eq:10}
\end{equation}
where the condition (\ref{eq:08}) for tangentially distorted images was inserted in the last step.

These simple considerations are summarised here because they enable us to draw several immediate conclusions on such galaxy clusters that reveal strong gravitational lensing. First, (\ref{eq:08}) shows that the mean scaled surface density inside a tangential critical curve must equal unity,
\begin{equation}
  \langle\kappa\rangle_\mathrm{t} =
  \frac{2\pi}{\pi\theta_\mathrm{t}^2}\int_0^{\theta_\mathrm{t}}\theta'\d\theta'\kappa(\theta') =
  \frac{m(\theta_\mathrm{t})}{\theta_\mathrm{t}^2} = 1
\label{eq:11}
\end{equation}
because of (\ref{eq:08}). If the mass profile decreases monotonically away from the cluster centre, as it can reasonably be assumed to do, $\kappa\gtrsim1$ must be satisfied for tangential arcs to occur. Strong lenses thus require dense cores. Thin tangential arcs require a radial magnification $\lesssim1$ at the tangential critical curve, or $\kappa\lesssim1/2$ according to (\ref{eq:10}). Since tangential arcs are typically thin, cluster density profiles have to be sufficiently steep that $\kappa$ can fall from $\kappa\gtrsim1$ near the core to $\kappa\approx1/2$ at the tangential critical curve. Radial arcs cannot be formed by lenses with very steep mass profiles, as the following rough estimate shows. Suppose $\kappa$ is constant near the lens centre, then the mass $m(\theta)$ grows $\propto\theta^2$. Then, the condition (\ref{eq:09}) is marginally satisfied. Detailed calculations with specific examples show that lenses with somewhat steeper density profiles can also produce radial arcs \citep{BA96.1}. Nonetheless, the general conclusion holds that the thin tangential arcs and the presence of radial arcs indicate that clusters have dense cores with fairly flat density profiles that steepen quickly. These are profound, qualitative conclusions quite independent of the detailed mass distribution in lensing clusters.

Arcs allow cluster masses to be easily estimated. The total, projected mass enclosed by a tangential critical curve is
\begin{equation}
  M(\theta_\mathrm{t}) = 2\pi\int_0^{r_\mathrm{t}}r\d r\Sigma(r)\;,
\label{eq:12}
\end{equation} 
where $r = D_\mathrm{l}\theta$ is the physical spanned by the angular radius $\theta$. Thus,
\begin{equation}
  M(\theta_\mathrm{t}) = \pi D_\mathrm{l}^2m(\theta_\mathrm{t})\Sigma_\mathrm{cr} =
  \pi D_\mathrm{l}^2\theta_\mathrm{t}^2\Sigma_\mathrm{cr} =
  \frac{c^2}{4G}\frac{D_\mathrm{l}D_\mathrm{s}}{D_\mathrm{ls}}\theta_\mathrm{t}^2
\label{eq:13}
\end{equation} 
where (\ref{eq:11}) and (\ref{eq:05}) were used. Such a mass estimate evidently requires the redshifts of the lens and the source to be known. If the cluster is very close, a fair approximation is
\begin{equation}
  M(\theta_\mathrm{t}) \approx \frac{c^2}{4G}D_\mathrm{l}\theta_\mathrm{t}^2\;.
\label{eq:14}
\end{equation} 

\subsection{Cluster masses}

It was soon discovered that the masses obtained this way from strong gravitational lensing are very close to mass estimates derived from X-ray observations. It is by no means obvious that this should be the case. Gravitational lensing is sensitive to the projected mass irrespective of its physical state, while the interpretation of X-ray data requires assumptions on symmetry and hydrostatic equilibrium of the gas with the gravitational potential well, if not even on isothermality of the intracluster gas. Even a qualitative, overall agreement between these entirely different mass estimates is thus a reassuring result. Nonetheless, a systematic discrepancy between detailed mass estimates was soon revealed in the sense that masses derived from strong lensing were typically larger by factors of $\simeq2\ldots3$ than masses obtained from X-ray observations \citep{WU94.2, MI95.2, WU96.1}. There are numerous more recent examples. Many find substantially discrepant mass estimates based on X-ray and strong-lensing observations \citep{PR05.1, GI07.1, MI08.1, HA08.1, EB09.1}, while good agreement is found in some other clusters \citep{RZ07.1, BR08.1, IS10.1}.

The reasons for discrepancies or agreement are not always straightforward to see. It is plausible, however, that good agreement is achieved in clusters which are relaxed and for which equilibrium assumptions can be assumed to hold, while unrelaxed clusters have a tendency to yield different mass estimates with different methods. 

On the whole, the following picture emerges. Massive clusters tend to have substantial non-thermal pressure support mainly due to gas turbulence and possibly also from magnetic fields \citep{KA10.1, MO10.2, MO11.2}. Bulk motion in the intracluster gas is seen in simulations \citep{2010A&A...519A..90M, 2010A&A...514A..93M, 2010A&A...519A..91F} of large cluster samples. This indicates that the gas in massive galaxy clusters is not fully thermalised yet, with part of the kinetic energy in the gas still being in ordered rather than unordered motion. In line with this argument, several authors used numerical simulations to show that X-ray mass estimates are generally biassed low \citep[see e.g.][Rasia06.1]. As discussed in \cite{BA96.2} this is particularly the case of merging clusters because their X-ray gas is still cooler than expected from their total mass, which is already seen by the lensing effect. This seems to explain the mass discrepancy at least in some clusters \citep[e.g.][]{SM95.1, OT98.1}. Incomplete or perturbed hydrostatic equilibrium of the intracluster gas can thus be expected to contribute to the observed discrepancy. Asymmetries and cluster substructures also play an important role. Due to their relatively larger shear, asymmetric and substructured clusters are more efficient lenses at a given mass. Mass estimates based on axially symmetric models are thus systematically too high \citep{BA95.1, HA98.1}.

The approximate triaxiality of galaxy-cluster haloes formed from cold dark matter was identified as a source of discrepant mass estimates even when hydrostatic equilibrium is assumed \citep{MO10.1, LI10.1, MO11.1}. Based on the assumption of hydrostatic equilibrium, axis ratios between $\sim1.25$ and $\sim2$ could be derived by combining X-ray and strong-lensing observations in a few individual galaxy clusters. Halo triaxiality also gives rise to an orientation bias, in the sense that the most efficient strong lenses tend to be prolate haloes whose longest axis is well aligned with the optical axis \citep{HE07.1, 2009MNRAS.392..930O, 2010A&A...519A..90M}. Another important indicator for dynamical activity of galaxy clusters is the offset between either the optical signal and the X-ray emission \citep{SC10.1} or between the baryonic matter and the dark matter reconstructed from gravitational lensing \citep{SH10.2}. Extreme cases of such offsets are clusters like the ``bullet cluster'' 1E~0657$-$558 \citep{CL06.1, CL04.1} where the X-ray emitting gas is located in between two clumps of dark matter identified by their lensing signal and the galaxies they contain, the ``cosmic train wreck'' Abell~520 \citep{MA07.2, JE12.1} where the dark matter coincides with the X-ray gas but not with the galaxies, and ``Pandora's box'' Abell~2744 \citep{2011MNRAS.417..333M} where three, possibly four, subclusters are seen in the process of merging. Numerical simulations indicate that such ``bullet clusters'' are not expected to be rare \citep{FO10.1}.

\cite{AL98.1} distinguished clusters with and without cooling flows and found an appreciable mass discrepancy in clusters without, but good agreement of X-ray and lensing mass estimates in clusters with cooling-flow. This supports the concept that well-relaxed clusters which had sufficient unperturbed time to develop a cooling flow are well-described by simple, axially-symmetric models for lensing and the X-ray emission, while dynamically more active clusters tend to give discrepant mass estimates; this was confirmed by \cite{WU00.1}. In a sample of 20 strongly lensing clusters, \cite{RI10.1} found that the strong-lensing masses are on average 30~\% larger than the X-ray masses at 3-$\sigma$ significance, with more substructured clusters showing larger discrepancies. \cite{MA99.1} noted that the mass discrepancy is reduced if cluster density profiles are steeper than inferred from the X-ray emission. It thus appears that discrepancies between X-ray and lensing masses can commonly be traced back to dynamical activity in unrelaxed clusters \citep[see also][]{SM05.1}, incomplete thermalisation of the intracluster gas, deviations from spherical symmetry as expected in cosmic structures formed from an initially Gaussian random field of cold dark matter and merging of subclusters. At least part of the disagreement occurs because of model restrictions which, if removed, generally lead to better agreement \citep{GA05.1, DO09.1}. It should also be kept in mind that strong gravitational lensing constrains the innermost core of a cluster's mass distribution. \cite{2011A&A...528A..73D} demonstrate, for example, that X-ray and strong-lensing mass estimates agree well in the core of Abell~611 where they are both well constrained, but tend to disagree farther away from the cluster centre where the strong-lensing mass estimates needs to be extrapolated \citep[see e.g.][for a study based on a large number of simulated cluster haloes]{2010A&A...514A..93M}. Selection effects play an important and not fully understood role in the interpretation of cluster lenses. While strongly-lensing clusters are more X-ray luminous than others, approximately 30--40~\% of the total abundance of strongly-lensed images are lost if only clusters with high X-ray luminosities are selected (Horesh et al. 2010, see later discussion).

\subsection{Cluster mass profiles}

Assuming mass profiles with cores, tangential arcs require small core radii as described above, but the existence of radial arcs requires finite cores of some sort \citep{LE94.1, LU99.1}. Despite difference in detail, numerical simulations of CDM halos unanimously show that density profiles flatten towards the core, but do not develop flat cores \citep{NA96.1, NA97.1}. \cite{BA96.1} showed that radial arcs can also be formed by halos with such ``cuspy'' density profiles, provided the central cusp is not too steep.

In principle, the relative abundances and positions of radial compared to tangential arcs in clusters provide important constraints on the central density profile in clusters \citep{MI95.1, 2001ApJ...559..544M, OG01.1}. Radial arcs are still too rare for successfully exploiting this method. Being much closer to the cluster cores than tangential arcs, they are also more likely to be confused with, or hidden by, the light of the cluster galaxies. Following \cite{MI95.1}, \cite{SA05.1} compiled a sample of clusters containing radial and tangential arcs and added constraints on the central mass profile from velocity-dispersion measurements in the central cluster galaxies. They demonstrated that, assuming axially-symmetric mass distributions, central density profiles have to be substantially flatter than those found in CDM simulations. However, even small deviations from axial symmetry can invalidate this conclusion and establish agreement between these observations and CDM density profiles \citep{BA04.1, ME07.1}. In a recent study by \cite{NE11.1}, a density profile somewhat flatter than expected in CDM was confirmed in the cluster Abell~383, but the tension seems to have diminished.

Attempts at modelling arcs with isothermal mass distributions are typically remarkably successful \citep[see][for an impressive early example]{KN96.1}. This is all the more surprising as numerical simulations consistently find density profiles which are flatter than isothermal within the scale radius and steeper outside. In a very detailed analysis, \cite{GA03.1} find that an isothermal core profile for the cluster MS~2137 is preferred compared to the flatter NFW profile. \cite{SM01.1} constrain the core density profile in A~383 using X-ray, weak-, and strong-lensing data and find it more peaked than the NFW profile, but argue that this may be due to the density profile of the cD galaxy. Similarly, \cite{KN03.1} find in a combined weak- and strong-lensing analysis of Cl~0024$+$1654 that an isothermal mass profile can be rejected, while the NFW profile fits the data well (note however that this cluster might have multiple mass components along the line of sight). \cite{SH08.2} show that the strong-lensing effects in two clusters A~370 and MS~2137 can be explained similarly well by isothermal and NFW density profiles, leading to substantial uncertainties in derived cluster properties and magnifications. Strong lensing alone constrains cluster density profiles only close to cluster centres, leaving considerable freedom in the mass models. It is also clear that baryons, in particular by cooling and star formation, can affect cluster density profiles where the gas density is high enough for cooling times to fall below the Hubble time. Certainly, the innermost cluster density profiles can be significantly influenced and steepened by baryonic physics \citep[see][for example]{BA10.2}.

\section{Arc abundances and statistics}
A different approach to study the mass distribution of strong lensing galaxy clusters makes use of statistical methods. The basic idea behind this approach is that the probability to observe strong lensing features depends on a number of lens properties which characterize their mass profile and their matter distribution. Last but not least, this probability is determined by the geometry of the universe. Since both lens properties and geometry of the universe depend on cosmology, the statistics of strong lensing features is potentially a tool for constraining the average structural properties of galaxy clusters and the cosmological parameters. Among the strong lensing features, this argument is  particularly applicable to the most elongated gravitational arcs, since these are rare, highly non-linear effects, whose appearance is extremely sensitive to the properties of the cluster cores. In the following sections, we will explicit the connection between statistics of the so called {\em giant arcs} and cosmology, and we will review what arc statistics studies taught us on the inner structure of strong lensing clusters. It is worth mentioning that the definition of giant arcs is still not well defined. \cite{WU93.1} used the term {\em giant arcs} to identify arcs with a length-to-width ratio exceeding ten and apparent $B$-magnitude less than $22.5$. This definition was used in the pioneer studies on arc statistics by \cite{BA98.2}. In many theoretical papers on arc statistics, the authors only use the length-to-width ratio exceeding a certain limit to classify an arc as {\em giant} \citep[e.g.][]{ME00.1,DA03.1,OG02.1}, being the limit $10$ or sometimes smaller \citep[see e.g.][who adopt a minimal length-to-width ratio of 7.5]{2010A&A...519A..90M}. The comparison between theoretical predictions and observations is complicated by the fact that a definition purely based on the length-to-width ratio does not take into account the detectability of arcs above the background noise in the real astronomical images. 

\subsection{Definitions}

In this Section, we introduce some basic definitions and formulate how the total number of images with a given property can be computed.

First, we define the lensing {\em cross section} $\sigma_Q$ for images
with property $Q$ to be the area in the source plane within which a source 
 has to lie in order to be imaged with property $Q$.  The cross
section is a function of some lens and source properties. We may
indicate these properties using a vector notation. In the following,
$\vec{p}_{\rm l}$ will indicate the properties characterizing the
lens. These should include the parameters defining the shape of the
density profile, the substructure content, the mass distribution, the
redshift, etc. Similarly, $\vec{p_{\rm s}}$ will indicate the properties  characterizing the sources, like for example their surface brightness distribution and size. Obviously, given the geometrical nature of lensing, the cross section depends on the redshift of both the lens and the sources, $z_{\rm L}$ and $z{\rm s}$. Using this notation, the lensing cross section for images with the property $Q$ will be given by
\begin{equation}
  \sigma_Q \equiv \sigma_Q(z_{\rm l}, z_{\rm s},\vec{p_{\rm l}},\vec{p_{\rm s}}) \ .
  \label{equation:sigmacr}
\end{equation}
Under the assumption that (1) cross sections do not overlap, (2)
source positions are uncorrelated with lens positions, and (3) that lens and source properties do not depend on redshift, the {\em
optical depth} $\tau$ is determined by adding up all cross section
attached to the lenses between the observer and the source, and
dividing the result by the area of the source plane. Therefore, we may write
\begin{equation}
	\label{eq:taugen}
	\tau_Q(z_{\rm s},\vec{p_{\ssmall}})=\frac{1}{4\pi D_S^2}\int_0^{z_{\ssmall}} \d z_{\dsmall} (1+z_{\dsmall})^3 \left| \frac{\d
  V(z_{\dsmall})}{\d z_{\dsmall}} \right| 
  \int_{\vec{p_{\dsmall}}}\d \vec{p_{\dsmall}} \frac{\d
  n(\vec{p_{\dsmall}},z_{\dsmall})}{\d \vec{p_{\dsmall}}}\sigma_Q(z_{\dsmall},z_{\ssmall},\vec{p_{\dsmall}}\vec{p_{\ssmall}})  \;.
\end{equation}
Here $\d V(z)$ is the cosmic volume element between redshift $z$ and $z+\d
z$, 
\begin{equation}
  \d V(z) = 4 \pi D^2(z)\frac{\d D_{pr}(z)}{\d z} \d z \ ,
\end{equation}
where $D_{pr}$ is the proper distance. The function $\d n(\vec{p}_{\dsmall},z_{\dsmall})/\d \vec{p}_{\dsmall}$ is the number density of lenses with properties $\vec{p}_{\dsmall}$ at redshift $z_{\dsmall}$.

The last integral in Eq.~\ref{eq:taugen} should be performed over the whole parameter space defining the lenses. This is obviously analytically impossible, unless simplistic assumptions can be made on $\vec{p}_{\dsmall}$. If we assume that only the mass characterizes the lensing cross section of the lens, then $\d n(\vec{p}_{\dsmall},z_{\dsmall})/\d \vec{p}_{\dsmall} = \d n(M,z)/\d M$ coincides with the mass function. If we further assume that the sources are all identical ($\vec{p}_{\ssmall}={\rm const}$), then the optical depth can be written as  
\begin{equation}
  \tau_Q(z_{\ssmall}) = \frac{1}{4 \pi D_{\ssmall}^2}
  \int_0^{z_{\ssmall}} \d z_{\dsmall} (1+z_{\dsmall})^3 \left| \frac{\d
  V(z_{\dsmall})}{\d z_{\dsmall}} \right| \int_0^{\infty} \d M \frac{\d
  n(M,z_{\dsmall})}{\d M} 
  \sigma_Q(M,z_{\dsmall},z_{\ssmall}) \ .
  \label{eq:optd}
\end{equation} 

Finally, the total number of images with property $Q$ is given by the
integral
\begin{equation}
  N_Q= \int_0^{\infty} \tau_Q(z_{\ssmall}) n_{\ssmall}(z_{\ssmall}) \d
  z_{\ssmall} \
  ,
  \label{eq:arcnum}
\end{equation}
where $n_{\ssmall}(z)$ is the number of sources at redshift $z$.

Looking at Eqs.~\ref{eq:taugen}, \ref{eq:optd}, and \ref{eq:arcnum}, it is clear that the number of arcs with property $Q$ depends on cosmology through:
\begin{itemize}
\item the geometry of the universe, here given by distances and volumes;
\item the structure formation, through the mass function and the sensitivity of internal properties of the lenses (like for example the concentration) to the value of cosmological parameters.   
\end{itemize}
All this makes {\em arc statistics} a potentially powerful cosmological tool. 

\subsection{Arc statistics as a cosmological tool}
\label{arcstatcosmo}

As briefly outlined in the previous section, we expect that arc
statistics is sensitive to cosmology. In more detail, the reasons for this
sensitivity are the following:
\begin{itemize}
\item gravitational arcs are images of background
  galaxies strongly distorted by massive and compact clusters;
\item typical sources for gravitational arcs are at redshifts
  $z_{\rm s} \gsim 1$. For background sources at these redshifts,
  clusters at $0.2 \lsim z_{\rm d} \lsim 0.4$ are the most efficient
  lenses;
\item if arcs are to be produced in abundance, a comparatively large
  number of massive and compact clusters must be in place at those
  redshifts. Galaxy
  clusters form earlier in low-density than in high-density models. Dark energy influences the
  growth of the density perturbations and the formation time of galaxy
  clusters as well. The number density of massive clusters in the
  redshift interval useful for strong lensing is thus expected to
  be highly sensitive to the content of the universe (in particular to the content of dark matter and dark energy);
\item the earlier the clusters form, the more concentrated they are,
  as indicated by numerical simulations of structure formation. The
  characteristic density of a numerically simulated dark matter halo
  indeed reflects the mean cosmic density at the time when the halo
  formed;
\item strong lensing is a highly non linear effect and the number of
  strong lensing events depends sensitively on the number of cusps in,
  and the length of, the caustic curves of the lenses. Cusps require
  asymmetric lenses. Asymmetric, substructured clusters are thus much
  more efficient in producing large arcs than symmetric clusters,
  provided that the individual cluster sublumps are compact
  enough. Clusters which are assembling from  compact subclusters at
  redshifts where the lenses are most efficient should thus
  produce many more arcs than clusters forming at lower redshifts from
  less compact sublumps;
\item finally, the volume per unit redshift is larger in low-density
  than in high-density cosmological models. Under the simple
  assumption of a constant number density of galaxy clusters, the
  number of possible efficient lenses between observers and distant
  sources is thus higher in low-density compared to high-density
  models.
\end{itemize} 

Indeed, the expected number of {\em giant\/} arcs,  changes by orders of
magnitude between low- and high-density universes according to the
numerical models described in \cite{BA98.2}. Their results are summarized in
Tab.~(1). 

\begin{table}
\label{tab:narcs} 
\begin{center} 
\begin{tabular}{lc} 
\hline\hline 
Cosm. Model & nr. of arcs\\ 
\hline\hline 
SCDM    &  36 \\ 
$\Lambda$CDM  &  280\\ 
OCDM  & 2400 \\ 
\hline\hline 
\end{tabular} 
\end{center} 
\caption{Number of giant arcs on the whole sky expected from the numerical
  study by Bartelmann et al. (1998). Simulations were carried out in
  different cold dark matter models: we report here the results for a
  flat model without cosmological constant (SCDM; $\Omega_0=1$,
  $\Omega_{0\Lambda}=0$, $h=0.5$), for a low-density flat model
  ($\Lambda$CDM; $\Omega_0=0.3$, $\Omega_{0\Lambda}=0.7$, $h=0.7$),
  and for a low-density open model (OCDM; $\Omega_0=0.3$,
  $\Omega_{0\Lambda}=0$, $h=0.7$).}
\end{table}

\subsection{Arc lengths and widths}
\label{sect:meth}
Giant arcs are gravitational arcs whose length-to-width ratio exceeds a certain threshold, which is by convention adopted to be $10$ \citep{WU93.1}. Since arcs with such large distortions are extremely rare, a lower limit of $7.5$ may be adopted in order to increase the number of images available for a statistical analysis. Independent on this issue, it is however not obvious how to define the length and the width of gravitational arcs. The following methods have been employed in theoretical studies, including ray-tracing techniques. We will assume to deal with pixelized images.
\begin{enumerate}
\item The method proposed by \cite{MI93.1} and
\cite{BA94.1} and later used in several other studies \citep[e.g.][]{ME01.1} consists of defining the area of
each image as the total number of image pixels. The perimeter is
defined as the number of boundary pixels.
 
It is then constructed a circle crossing three image points, namely (1) its
center, (2) the most distant boundary point from (1), and (3) the most
distant boundary point from (2). The center of the image is given by the pixel which is mapped on the source plane nearest to the source center. Notice that long
arcs can be merged from several images, and there may exist more than
one image of the source center. However, this is not a problem because
these points are located almost on the same circle. Thus, it is adopted
one of them as representative of the image center.

\begin{figure}[t]
\begin{center}
\resizebox{4cm}{!}{\includegraphics{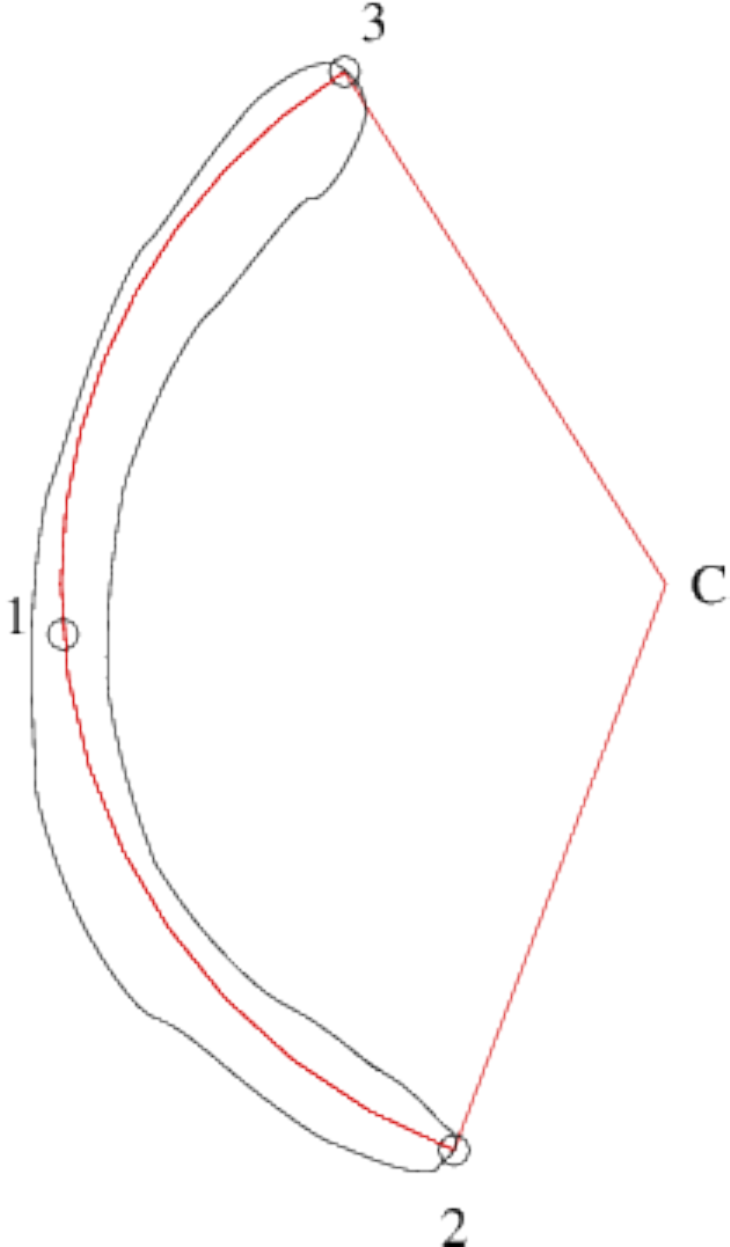}} 
\end{center}
\caption{Procedure for measuring the arc length. Three arc points are used: (1) the image of the source
  center, (2) the most distant boundary point from (1) and (3) the
  most distant boundary point from (2). The length of the circle segment passing through these three points are
  assigned to the arc.}
\label{figure:arcfit}
\end{figure}
 
The length $l$ of the image is defined 
through the circle segment connecting points (2) and (3)
(Fig.~\ref{figure:arcfit}). For determining the image width $w$,  a simple geometrical figure with equal area and length is searched. For
this fitting procedure,  ellipses, circles, rectangles and
rings are considered. In the various cases, the image width is approximated by the
minor axis of the ellipse, the radius of the circle, the smaller side
of the rectangle, or the width of the ring, respectively. A possible
test for the quality of the geometrical fit is given by the agreement
between the perimeter of the geometrical figure and the image.
\item \cite{DA03.1} and \cite{HE07.1} use the same method outlined above to measure the arc length, while the arc width is derived from the image area using the relation
\begin{equation}
w=A/l \;.
\end{equation}
This implies that the definition of $l/w$ is formally identical to that of the previous method, if all images are well fitted by rectangles.   
\item \cite{WA03.1} approximates the  length-to-width ratio to the total magnification of the image. They claim that highly magnified sources are typically elongated in the tangential direction and they assume that the radial magnification is close to unity. This approximation holds strictly for lenses with isothermal density profile. One can generalize this definition to take into account radial magnification by setting the length-to-width ratio equal to the ratio between the eigenvalues of the Jacobian matrix (Eq.~\ref{eq:06}). Following this approach, the length-to-width ratio is given by 
\begin{equation}
 	l/w=\lambda_r/\lambda_t \;,
	\label{eq:eigrat}
\end{equation}
where $\lambda_r$ and $\lambda_t$ are the radial and the tangential eigenvalues, whose inverse are used to estimate the local radial and tangential magnifications of the image.
\end{enumerate}

By analyzing a sample of $\sim 700000$ lensed images obtained from ray-tracing simulations, we find that the vast majority of them are best fitted by ellipses ($\sim 44\%$) or rectangles ($\sim 55\%$). Only $\sim1\%$ of the images are well described by rings and the fraction of images similar to circles is negligible. Imposing a minimal length-to-width ratio of 5, the fraction of ellipses drops to $\sim 5\%$, and it further reduces to $1.6\%$ and $0.8\%$ for $l/w_{\rm min}=7.5$ and  $l/w_{\rm min}=10$, respectively. Conversely, the fraction of rectangle-like arcs is $\sim 99\%$ for the most elongated arcs. Therefore, the above methods 1 and 2 are equivalent, if applied to arcs with large $l/w$, while method 2 tends to under-estimate the the width by a factor 
\begin{equation}
Q_{w}=\frac{w_{\rm rectangle}}{w_{\rm ellipse}}=\frac{A}{l}\frac{l\pi}{4A}=\frac{\pi}{4}
\label{eq:corrw}
\end{equation}
for an increasing number of images if $l/w_{\rm min}$ is reduced.

\begin{figure}[t]
\begin{center}
\resizebox{12cm}{!}{\includegraphics{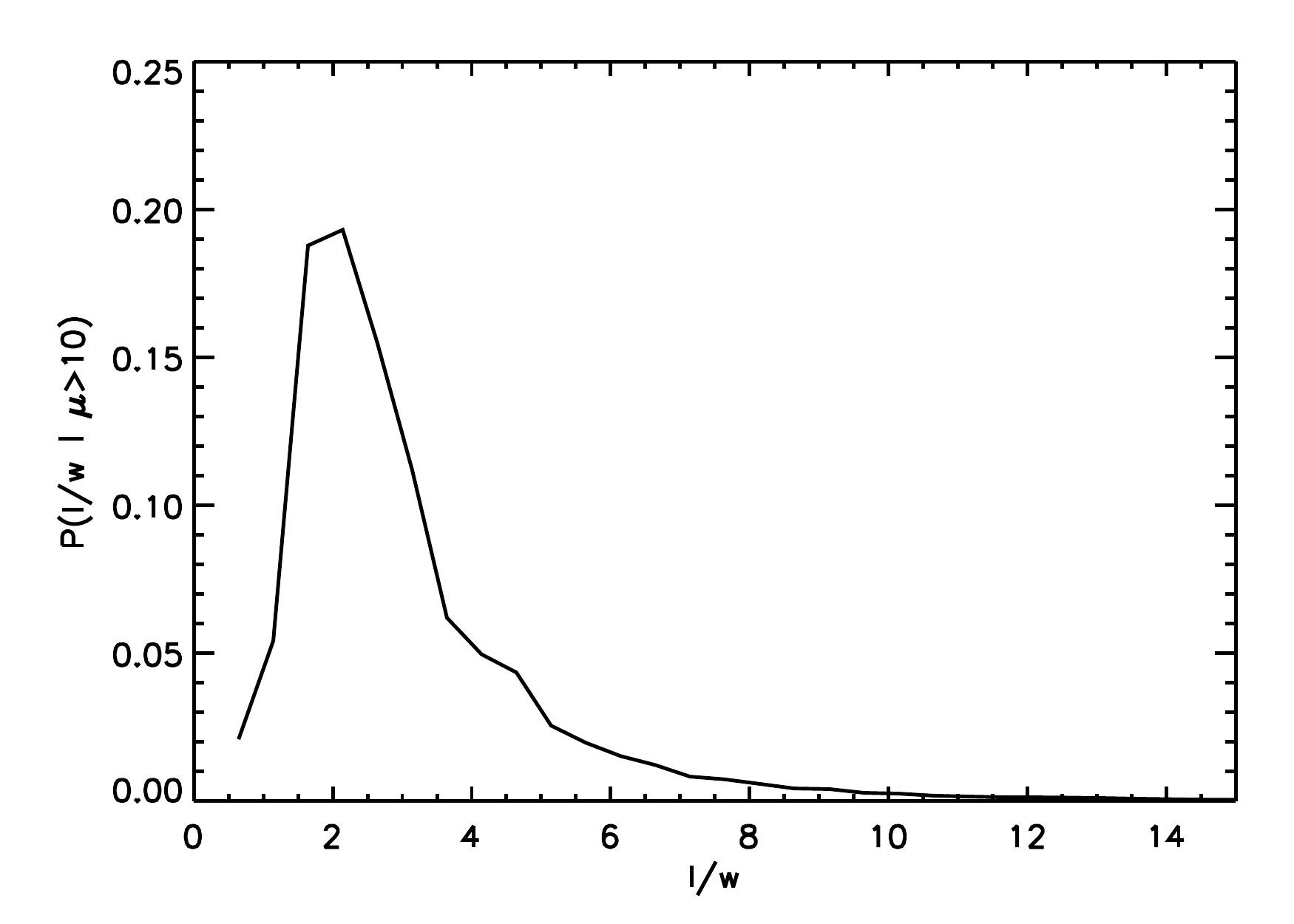}} 
\end{center}
\caption{Conditional probability distribution for the length-to-width ratio of arcs given a magnification $\mu>10$ obtained from a set of simulated arcs.}
\label{fig:lwmag}
\end{figure}

\begin{figure}[t]
\begin{center}
\resizebox{12cm}{!}{\includegraphics{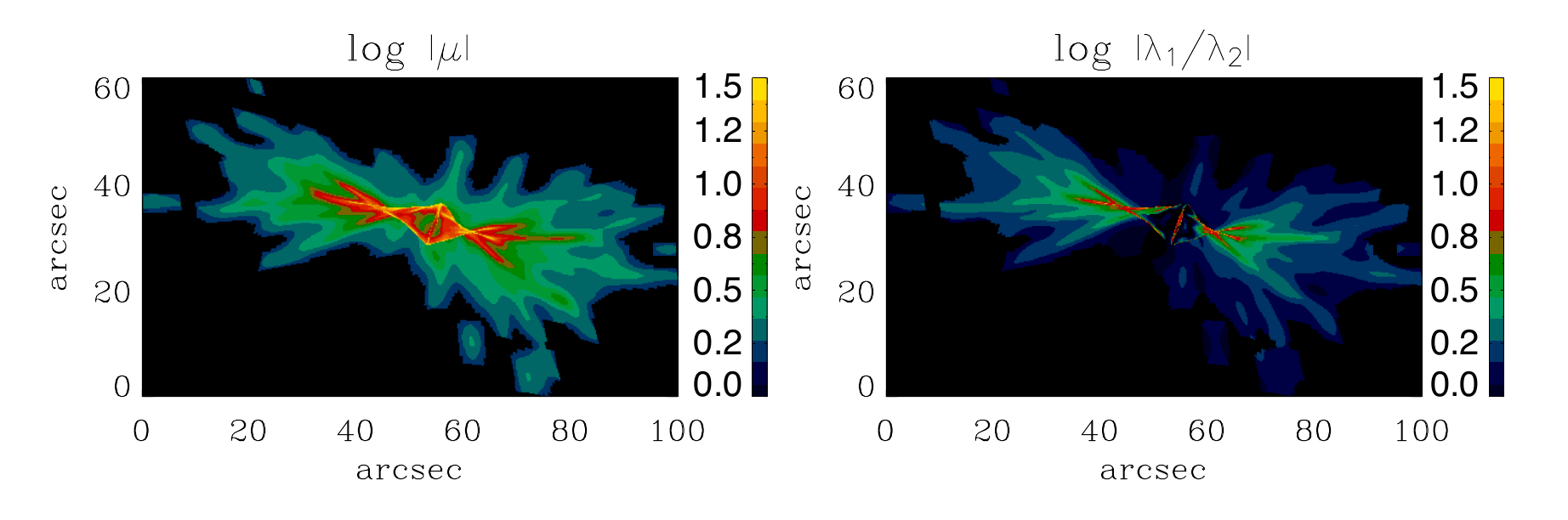}} 
\end{center}
\caption{Maps of absolute magnification (left panel) and eigenvalue ratios (right panel) for a dark matter halo taken from N-body simulations. Figure from \cite{LI05.1}}.
\label{fig:li2005fig1}
\end{figure}

In Fig.\ref{fig:lwmag} we show the conditional probability distribution for the length-to-width ratio of arcs given a magnification $\mu>10$. The distributions peak at $l/w\sim 2$, which indicates that images selected by magnification are generally characterized by small $l/w$ ratios. The reason is clear from Fig.~\ref{fig:li2005fig1}, taken from \cite{LI05.1}, which shows the maps of absolute magnification and of tangential-to-radial magnification ratio on the source plane of a numerically simulated galaxy cluster (left and right panels, respectively). In the left panel, we see that the high magnification region surrounds the whole caustics. In particular, there are source positions near both the radial and the tangential caustic where the magnification is very high, but the images are distorted both radially and tangentially, thus they do not appear as giant arcs. An observational example of such highly magnified sources is the spiral galaxy observed near the core of        
the cluster MACS J1149.5+2223 \citep{2009ApJ...703L.132Z,2009ApJ...707L.163S}. On the contrary, large tangential-to-radial magnification ratios, expressed by the ratio between the radial and the tangential eigenvalues of the Jacobian matrix, are reached by sources located near the cusps of the tangential caustic.

Using Eq.~\ref{eq:eigrat} to estimate the length-to-width ratio of an arc implies using local measurements of the eigenvalues $\lambda_r$ and $\lambda_t$. However, $\lambda_t$ is zero for sources exactly on the tangential caustics, which causes $l/w$ to diverge for such sources. Additionally, a source is not point-like, but rather extended. Indeed, $l/w$ should be evaluated by convolving $\lambda_r/\lambda_t$ with a function which differs from zero only on the region occupied by the source. \cite{FE05.1} developed a fast semi-analytic method to compute the arc cross sections which makes use of this convolution assuming that all sources are elliptical with random orientation and axial-ratio. Although they found a good agreement between their method and ray-tracing simulations implementing the above method 1 to measure the arc width, recently \cite{2012A&A...547A..66R} pointed out that the length-to-width ratios computed "a la Fedeli et al.'' need to be corrected by a constant factor. The reason is that this last method implicitly assumes that the resulting arcs differ from their sources only by mean of an extra-ellipticity. Thus, their $l/w$ is similar to that should be obtained with method 1, fitting all arcs with ellipses. Since most of the giant arcs are best fitted by rectangles in method 1, the correction factor to be applied to the $l/w$ ratios from method 3 to match method 1 (and 2) is given again in Eq.~\ref{eq:corrw}.
            
\section{The arc statistics problem}
\subsection{Origin of the problem}
The results of \cite{BA98.2}, reported in Tab.~(1), immediately resulted to be  puzzling. \cite{LE94.1} detected six arcs in a sample of 16 clusters selected for their high X-ray luminosity as measured by the EMSS satellite and noticed that  such incidence of gravitational arcs  is noticeably high. They argued that even more concentrated mass profiles than those used by \cite{BA95.2} are necessary for explaining them quantitatively. Comparing their results with the data from \cite{LE94.1}, \cite{BA98.2} concluded that only their cluster sample taken from a simulation with low matter density ($\Omega_0=0.3$) and no cosmological constant could well reproduce the measured high arc abundance, but the other tested models failed badly. In particular, a flat cosmological model with $\Omega_0=0.3$ and $\Omega_\Lambda=0.7$ produced an order of magnitude less arcs than observed. This inconsistency between expectations in the $\Lambda$CDM cosmological model, currently supported by a number of powerful cosmological probes, and the observed occurrence of gravitational arcs was lated named {\em the arc-statistics problem}.

\subsection{Proposed solutions}

\subsubsection{Complexity of strong lensing clusters}
\label{sect:complexsl}

The arc-statistics problem was disputed based on calculations using analytic models for cluster lenses \citep{CO99.2, KA00.1}, which failed to reproduce the strong dependence on the cosmological constant claimed by \cite{BA98.2}. The possible influence of cluster galaxies on the arc-formation efficiency of cluster lenses was investigated by \cite{FL00.1} and \cite{ME00.1}, but found to be negligible. Even though galaxies in clusters locally wind caustics and critical curves and thus increase their length, they also cause long arcs to split. \cite{MO99.1} confirmed that axially-symmetric mass models adapted to the X-ray emission do not produce a sufficient number of arcs. They found that using NFW profiles for the dark-matter profile helped, but the profiles required too high masses, and proposed that substructured mass distributions could be the solution. As already mentioned above, \cite{ME03.2} adapted elliptically distorted lenses with NFW mass profile \citep[see also][]{GO02.1} to numerically simulated clusters and found the analytic models inadequate for quantitative arc statistics despite the asymmetry, demonstrating the importance of substructures.

\cite{BA98.2} used the ray-tracing technique for
studying gravitational lensing by galaxy cluster models taken from
N-body simulations \cite[see also][]{BA94.1,BA95.1,BA95.2,ME00.1,ME01.1}. This allows the most realistic
description of the cluster lenses because all effects which could play
an important role for the lensing phenomena are by construction taken
into account. 

A number of studies have shown that simple analytical lens models fail to reproduce the cross section for giant arcs of cluster-sized halos obtained from numerical simulations. For example, \cite{BA94.1} used a numerically simulated galaxy cluster to show that asymmetric, substructured cluster models are significantly more efficient strong lenses than axially-symmetric mass distributions because of their enhanced tidal field. Averaging over a sample of simulated clusters, \cite{BA95.2} quantified that the cross sections for arc formation could be up to two orders of magnitude larger for asymmetric than for axially symmetric cluster models of the same mass. These results were confirmed by \cite{ME03.1}, who showed that even elliptically distorting the lensing potential of the lenses does not help simple and smooth mass distributions to become lenses as strong as  numerically simulated cluster halos. 

\cite{ME07.1} analyzed a sample of simulated halos and quantified separately the effects of ellipticity, asymmetries, and substructures on their cross sections for giant arcs. Their experiment consists of creating two-dimensional smoothed, differently elliptical
and asymmetric versions of the numerical models. By subtracting these
smoothed mass distributions from the corresponding numerical maps and by
gradually smoothing the residuals, before re-adding them to the clusters, they obtained several representations of the lenses, which differ for the amount of ellipticity, asymmetries, and substructures, but which have the save average surface density profile. They found that the individual contributions of the above mentioned lens properties amount to $\sim 40\%$, $\sim 10\%$ and $\sim
30\%$ of the total strong lensing cross section, respectively. 

\cite{OG03.1} studied the strong-lensing properties of triaxial (rather than ellipsoidal) halos and found that they may well explain the high arc abundance, provided their central density slopes are steep enough, with a double-logarithmic slope near $-1.5$. 

\cite{WI99.1} noted that the arc radii in clusters depend only weakly on clusters mass and suggested that massive cD galaxies may be the reason. However, \cite{ME03.1} studied the effect of cD galaxies on the overall arc abundance and found it insufficient to remove the arc statistics problem. If the cosmological constant is replaced by some form of dynamical dark energy, structures tend to form earlier during cosmic history. Since cluster core densities reflect the mean cosmic density at their formation time, clusters thus tend to be more concentrated in dark-energy compared to cosmological-constant models. \cite{BA03.1} estimated the effect of higher cluster concentrations on arc statistics by analytic means. They found that dark energy may in fact increase arc abundances noticeably, but again not sufficiently for solving the arc statistics problem.

Galaxy clusters at high redshifts are found to be remarkably efficient lenses \citep{GL03.1, ZA03.1, SC10.1} even though they should be by far not massive enough for producing large arcs. A particularly impressive example is the cluster RX~J105343$+$5735 at $z=1.263$ which contains a large arc from a source at $z=2.577$ \citep{TH01.1}. More recently, \cite{2012ApJ...753..163G} reported the discovery of a giant arc, probably originated by a source at $z\gtrsim 3$, in the cluster IDCS J1426.5+3508 at $z=1.75$. Semi-analytical calculations in the framework of a $\Lambda$CDM cosmology, suggest that this lens system should not exist.  

So far, all effects studied, including baryon cooling in cluster cores \citep{PU05.1, WA08.1} and line-of-sight projection effects \citep{PU09.1}, returned moderate enhancements of the expected arc abundance. This is particularly true in state-of-the-art simulations which include AGN feedback. For example, \cite{2012MNRAS.427..533K} showed that, while gas cooling and star formation alone increase the number of expected giant arcs, particularly for lower redshift clusters and lower source redshifts, the inclusion of AGN feedback brings the strong lensing cross sections back to values very similar to those measured in dark-matter only simulations.  
 
\subsubsection{Source properties} 
\label{sect:srednorm}
\cite{DA04.1} used numerical cluster simulations to estimate arc cross sections and found reasonable agreement with the earlier results of \cite{BA98.2}, but arrived at a higher expected arc abundance because they inserted a higher normalisation for the number density of both X-ray clusters and background sources.

\cite{WA03.1} simulated the magnification probability for light rays propagating through a section of the Universe and found that the abundance of high-magnification events depends strongly on the source redshift. They attributed this to the exponential mass function of massive halos, which leads to a steep increase with source redshift in the number of halos suitable for strong lensing. Identifying the probability for highly magnified light bundles on random patches of the sky with the probability for finding arcs in massive galaxy clusters, they suggested this result as the resolution for the arc statistics problem. 
   
Again using high resolution N-body simulations and closely following the approach of \cite{BA98.2}, \cite{LI05.1} found that the optical depth scales as the source redshift approximately as
\begin{equation}
\tau=\frac{2.25\times10^{-6}}{1+(z_{\rm s}/3.14)^{-3.42}}
\label{eq:tauli}
\end{equation}
for sources with diameter 1''. The scaling amplitude is $\sim 50\%$ larger for smaller sources with a diameter of 0.5''. Qualitatively, they confirmed the high sensitivity of the optical depth to the source redshift, but they found that 1) the amplitude of the optical depth is a factor of $\sim 10$ smaller  and 2) the optical depth increases slower with the  source redshift than found by \cite{WA03.1}. This is due to the choice of \cite{WA03.1} to approximate the $l/w$ ratio with the total magnification (see discussion in Sect.~\ref{sect:meth}).

The impact of several source properties on the efficiency of numerically simulated clusters to produce giant arcs was extensively discussed also in the paper by \cite{2009ApJ...707..472G}. Their analysis was focussed on a sample of ten simulated massive halos at redshift $0.2$ and $0.3$ taken from a $\Lambda$CDM cosmological simulation. They used the $I$-band data of HST/ACS in the COSMOS field to quantify the distributions of background galaxy sizes, shapes, redshifts, and clustering down to a limiting surface brightness of $25$ mag/arcsec$^2$. They found that the source size and clustering only have small effects on the production of giant arcs. They find that the number of giant arcs is {\em decreased} by a factor of 1.05 and 1.61 when the COSMOS redshift distribution of galaxies is used instead of placing all sources on single planes at redshift $z_{\rm l}=1$ and  $z_{\rm l}=1.5$, respectively. Additionally, they noted that $\sim 30\%$ of galaxies with very elongated shapes $e=1-b/a>0.5$ increases the production of giant arcs by a factor of 2, compared to simulations where the galaxy ellipticities were uniformly distributed in the range $0<e<0.5$.

\subsubsection{Power spectrum normalization}

The optical depth in Eq.~\ref{eq:tauli} is only 10\% of the optical depth measured by \cite{BA98.2} for $z_{\rm s}=1$. The reason was attributed by \cite{LI05.1} to the different normalization of the power spectrum of the primordial density perturbations, $\sigma_8$, used to generate their numerical halos. While a fraction of the halos used by \cite{BA98.2} came from a cosmological simulation with $\sigma_8=1.12$, \cite{LI05.1} used halos extracted from a cosmological box evolved with $\sigma_8=0.9$. In a follow-up paper, \cite{2006MNRAS.372L..73L} showed that the optical depth decreases by a factor of 6 changing from a cosmology with $\sigma_8=0.9, \Omega_m=0.3$ to a WMAP3 normalized cosmology with $\sigma_8=0.74$ and $\Omega_m=0.238$. The value of $\sigma_8$ was found to affect significantly the lensing optical depth also by \cite{2008A&A...486...35F}. 
The main problem is now that recent measurements converge on a low normalisation parameter $\sigma_8\approx0.8$ for the dark-matter power spectrum, which, as said,  drastically lowers the expected arc abundance. Thus, even assuming that the strong-lensing properties of individual clusters are sufficiently understood \citep{HO05.1, 2011MNRAS.418...54H}, the arc-statistics problem seems to persist.

\subsubsection{Selection effects}
Besides improved methods for predicting arc abundances \citep{FE05.1}, selection biases in existing samples of strongly lensing clusters must be understood before further progress will be made \citep{FE07.1}. Considering the statistics of gravitational arcs in a sample of 97 galaxy clusters observed with HST \cite{2010MNRAS.406.1318H} found that X-ray-selected clusters with redshifts $0.3\le z\le 0.7$ have significantly more arcs than optically-selected clusters with similar optical richness. While the X-ray selected clusters in their sample show $1.2\pm0.2$ arcs per cluster, optically selected clusters have only $0.2\pm0.1$ arcs per cluster. 

Some progress has been made on the theoretical side by investigating the properties of large samples of strong lensing clusters in numerical simulations. \cite{2010A&A...519A..90M} found that, at fixed virial mass, the strongest lenses have X-ray luminosities that are typically higher than the average. This bias is larger for small-mass  and high-redshift halos. Through a comparison with the cluster dynamical state, they showed that a large fraction of X-ray luminous systems are dynamically active systems, which  are far from virial equilibrium \citep[see also][]{2004MNRAS.352..508R}. 

\begin{figure}[t]
\begin{center}
\resizebox{12cm}{!}{
\includegraphics{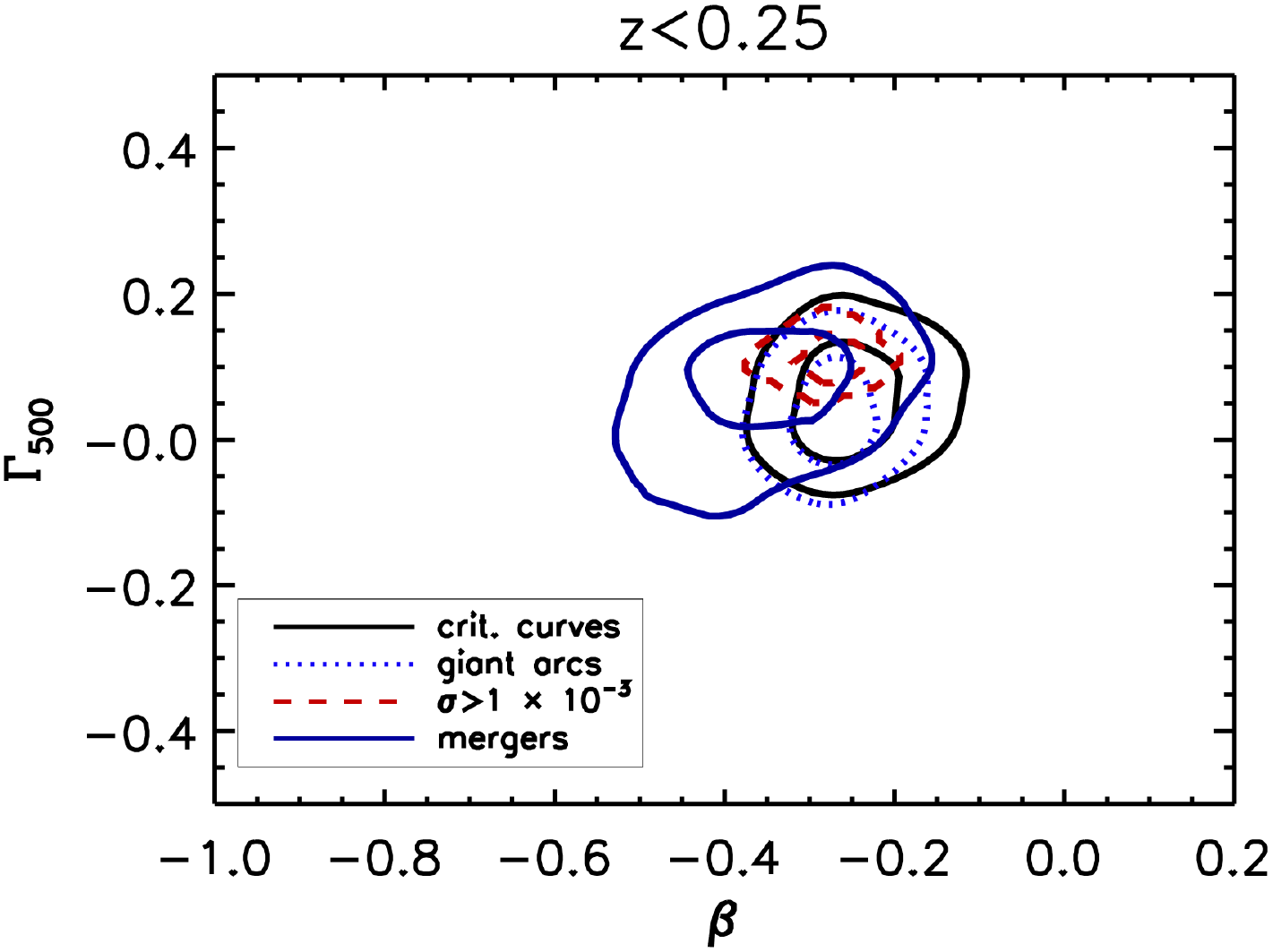}
\includegraphics{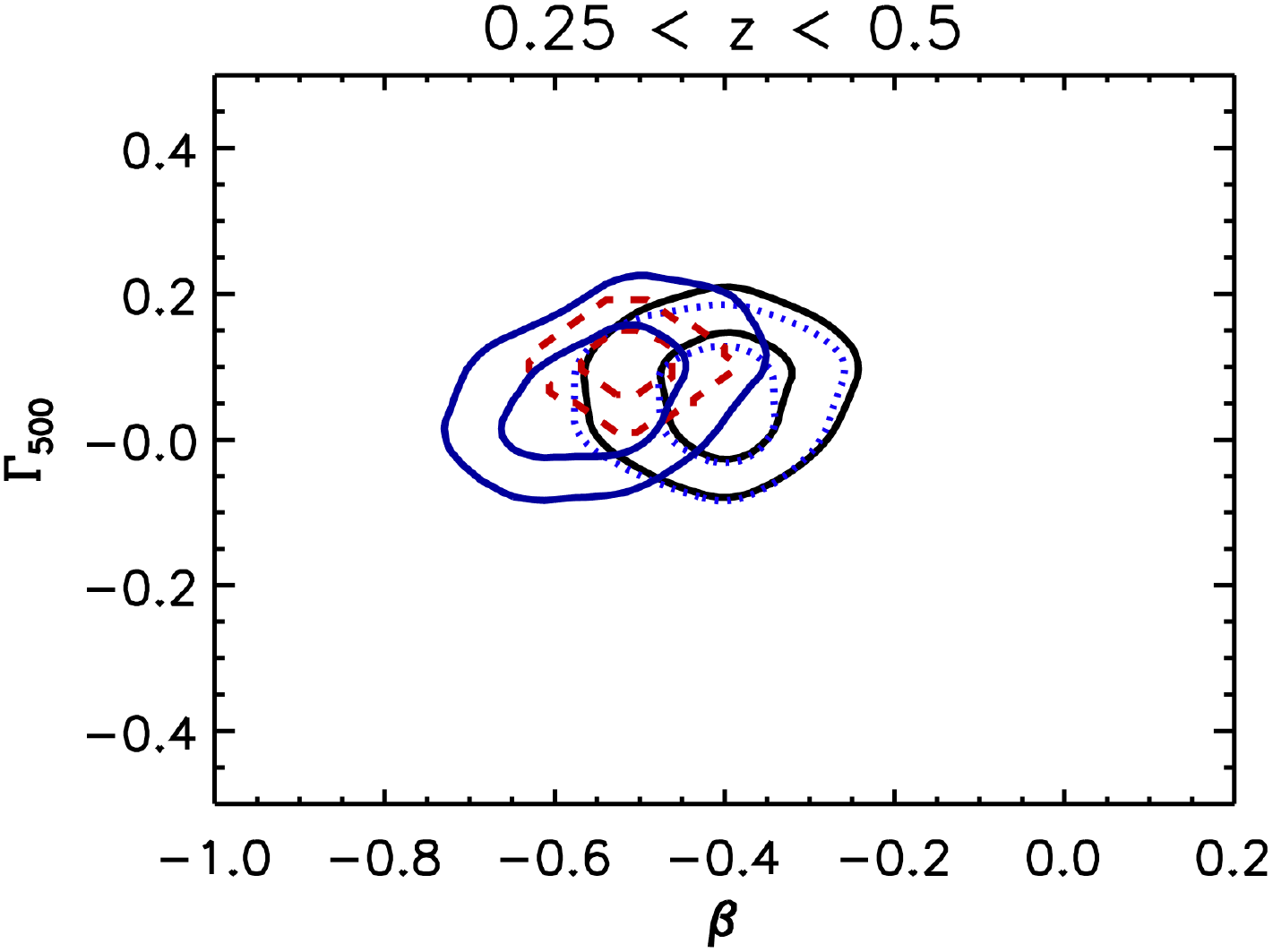}}
\resizebox{12cm}{!}{
\includegraphics{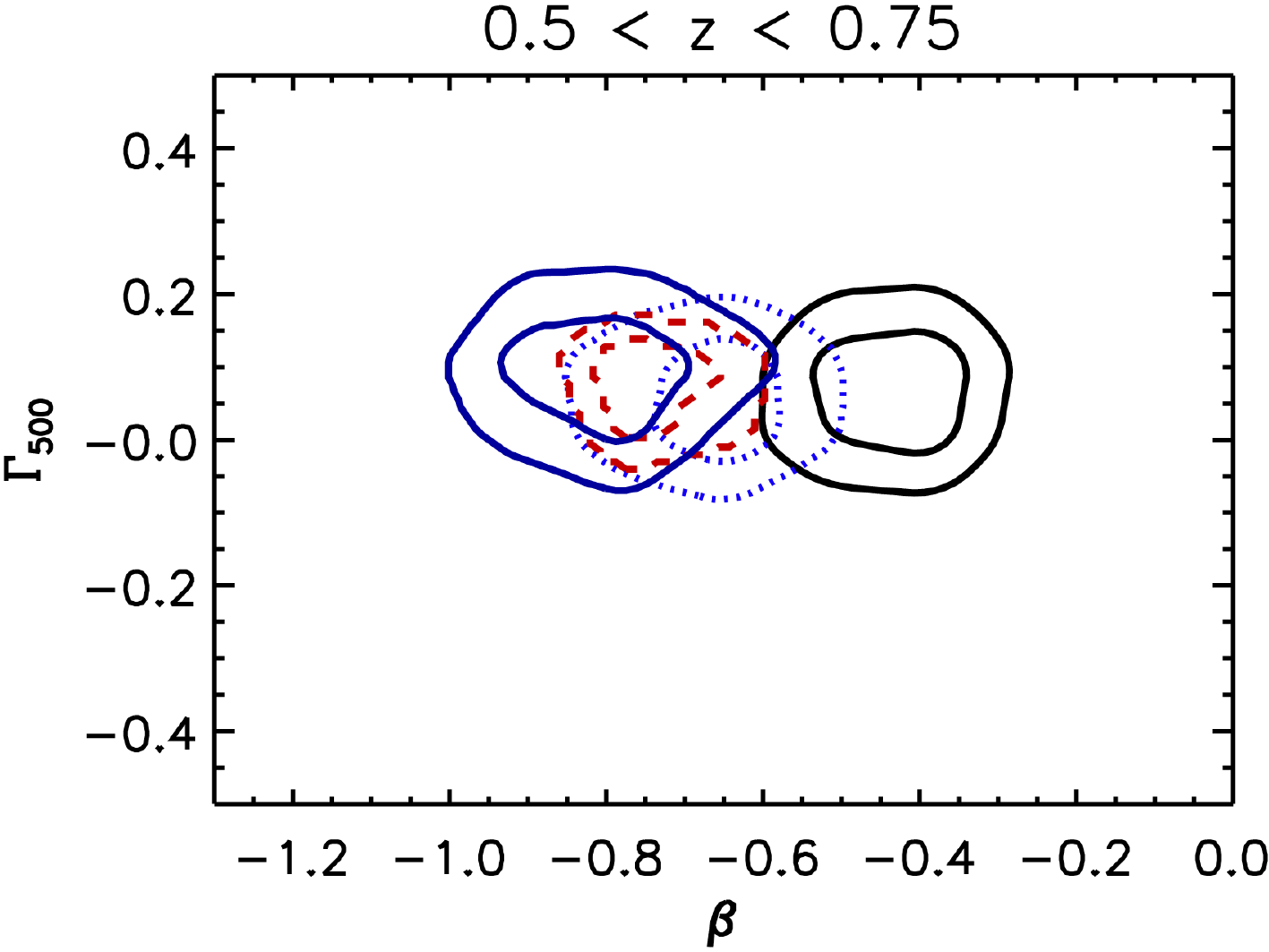}
\includegraphics{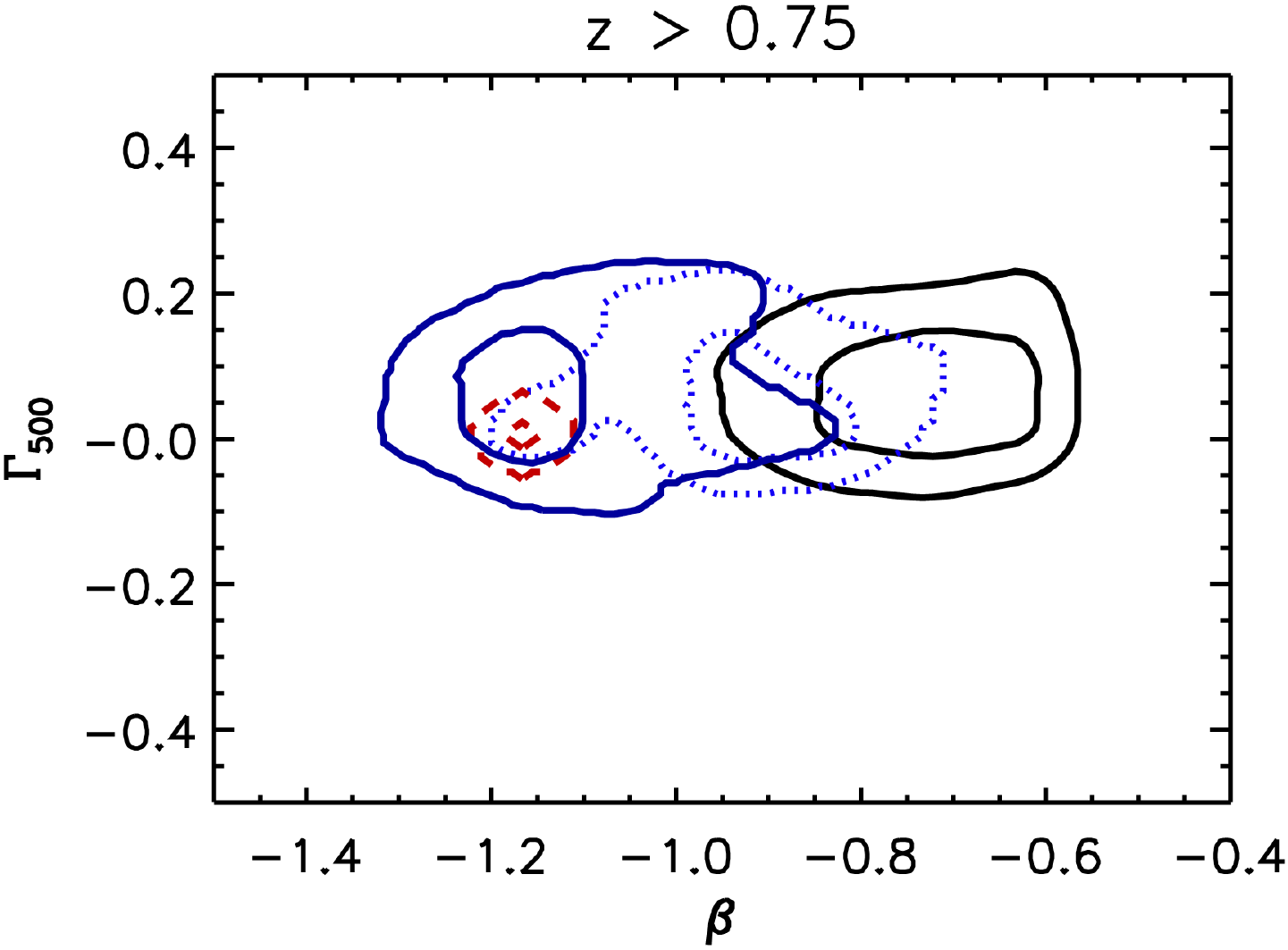}
} 
\end{center}

\caption{The distributions of lensing clusters in the cosmological hydrodynamical simulation {\sc MareNostrum Universe} \citep{2007ApJ...664..117G} in the $\beta-\Gamma_{500}$ plane. Results for clusters with increasing ability to produce large distortions are shown. The contours indicate the levels corresponding to 90\% and 50\% of the distribution peaks. The black solid contours show the distribution of all clusters that have critical lines for sources at $z_{\rm s} = 2$. The blue-dotted and the red-dashed contours show the distributions of clusters with minimal lensing cross sections for giant arcs $\sigma = 0$ and $\sigma = 10^{-3} h^{-2}$ Mpc$^2$, respectively. Finally, the blue-solid contours indicate the distribution of clusters identified as ``mergers". Figure from  \cite{2010A&A...519A..90M}.}
\label{fig:mergerslcorr}
\end{figure}

The correlation between strong lensing strength and dynamical state becomes particularly strong at high redshift. Let us quantify the departure from virial equilibrium using the parameter 
\begin{equation}	
	\beta=1+\frac{2T-S}{U} \;,   
\end{equation}
where $T$ and $U$ are the kinetic and the potential energies of the system, and $S$ is a surface pressure term that arises from considering the structure as contained in a limited region. Let us quantify the departure from hydrostatic equilibrium by means of the parameter 
\begin{equation}
\Gamma_{500}=1-\frac{M_{\rm hydro,500}}{M_{500}} \;,  
\end{equation}
where $M_{\rm hydro, 500}$ and $M_{500}$ are respectively the hydrostatic mass (i.e. recovered from the gas density and temperature assuming hydrostatic equilibrium, see the companion review by Ettori et al.) and the total mass of the system at the radius enclosing an over density of $500$ times the critical density of the universe. Numerical simulations show that the strongest lenses in a cosmological hydrodynamical simulation tend to populate the same region in the $\beta-\Gamma_{500}$ space which is populated by halos experiencing a merger phase. The correlation between strong lensing efficiency and merger activity is tighter at high redshift. This is shown in Fig.~\ref{fig:mergerslcorr}.

The fact that strong-lensing efficiency of clusters can be increased substantially and on a short timescale during a major merger was shown  with the help of numerical simulations by \cite{TO04.1}. As a subcluster approaches a cluster, the tidal field is increased, leading to a first maximum of the cross section approximately when the two clumps are few hundreds kpc distant. The cross section then slightly decreases and approaches a second maximum when the separation of cluster and subcluster is minimal. A third peak corresponding to the first is formed when the subcluster leaves again after the merger. During that process, the arc cross section can change by an order of magnitude or more on a time scale of $\simeq0.1\,\mathrm{Gyr}$. It thus appears that strong cluster lensing may be a transient phenomenon at least in some clusters which would otherwise be not massive or concentrated enough. The dependence of the main merger epoch on cosmic history would then establish an interesting link between high-redshift, strong cluster lenses and the cosmological framework model. Using semi-analytic methods, \cite{FE05.1} estimated that mergers approximately double the strong lensing optical depth for lens redshifts $z_{\rm L}>0.5$ and source redshifts $z_{\rm s}\sim2$.

An additional selection bias affecting the strong lensing clusters may be due to their three-dimensional shapes. As discussed in \cite{2012arXiv1210.3067L}, clusters are supposed to have triaxial shapes, being preferentially described by prolate ellipsoids. This picture, which is supported by numerical simulations, indicates that the strongest lensing effects are observed when the lenses are accidentally oriented such that their major axes point towards the observer. Under these circumstances, the clusters may appear relatively round also in the X-ray. Clusters are frequently classified as relaxed on the basis of X-ray morphological indicators \citep[see e.g.][]{2010ApJ...721L..82C}. Thus, X-ray selected samples of {\em relaxed} clusters may include a large fraction of clusters elongated along the line of sight. \cite{HE07.1} and \cite{2010A&A...519A..90M} showed that numerically simulated strong lenses extracted from cosmological boxes are affected by an orientation bias, and that the bias is stronger for the halos with the largest cross sections.     
   
\subsubsection{Alternative Cosmologies}

Could the arc statistics problem be cured by modifying the cosmological model? This is plausible because cosmology enters in two ways into the strong-lensing properties of galaxy clusters. First, the cosmological expansion function determines the volume and the lensing efficiency through the angular-diameter distance. Second, clusters may form earlier or later in other cosmological models than $\Lambda$CDM, affecting their internal matter concentrations as well as the evolution of their number density with redshift.

Using an analytic lens model based on the NFW density profile, developed and tested previously by \cite{ME03.1}, the effect of dark energy with a constant equation-of-state parameter $w\ne-1$ on the optical depth for the formation of giant arcs was studied by \cite{2003A&A...409..449B}. They showed that even substantial changes in $w$, well beyond current observational constraints, could only be expected to change the optical depth by at most a factor of two. This was later confirmed by \cite{2005A&A...442..413M} in a fully numerical study of several cosmological models whose $w$ parameter varied with time according to a variety of models for dynamical dark energy. The main result was that the dark-energy models studied interpolate between the standard $\Lambda$CDM model and spatially open CDM models with the same low matter density but without cosmological constant. The cluster concentration was identified there as the decisive parameter for modified strong-lensing efficiency. The paper concluded writing that constraints on dark energy from arc statistics were only possible if the normalisation parameter $\sigma_8$ was precisely known.

A particular class of cosmological models with an appreciable dark-energy density at high redshift, so-called early dark-energy models, was semi-analytically tested for their arc abundance by \cite{2007A&A...461...49F}. While they found that early dark energy increases the optical depth for giant arcs in particular at cluster redshifts $\gtrsim1$, this result was questioned later when it was realised that the $\delta_\mathrm{c}$ parameter entering the mass function had been calculated based on inapplicable assumptions.

While cosmology affects arc statistics through geometry as well as the dynamics of structure formation, purely geometrical constraints can be obtained from individual clusters containing multiple arc systems at different redshifts. Sources at different redshifts are lensed by the same mass distribution, whose lensing efficiency increases with the distance from the lens to the source. From the ratio of the measureable lensing efficiencies, the expansion function can be constrained, and thus the cosmological model. For this to work, the lensing mass distribution needs to be known sufficiently well, which causes degeneracies between the mass model and the geometrical constraints on cosmology.

This technique was applied to the massive cluster Abell 1689 by \cite{2010Sci...329..924J}, who described the cluster mass distribution by a detailed parametric model. They showed that the dark-energy equation-of-state parameter could be constrained to $w = -0.97\pm0.07$ in this way if their results were combined with constraints from the CMB through WMAP-5 and from the abundance of X-ray clusters. Regarding the same geometrical technique, \cite{2011MNRAS.411.1628D} noted that individual clusters give poor constraints, but competitive results may be obtained from a sample of about ten clusters containing as many as 20 arc families each.

Modifications of the cosmological model could also concern the initial conditions of structure growth. In this context, \cite{2011MNRAS.415.1913D} studied how non-Gaussian initial density fluctuations might modify the abundance and the internal concentration of galaxy cluster-sized lenses. Accepting the fairly tight limits on the lowest-order non-Gaussianiy parameter $f_\mathrm{NL}$ obtained from WMAP-7, they found that the arc optical depth for sources at $z_\mathrm{s}\approx2$ is enhanced by $\approx20\,\%$ for the $f_\mathrm{NL} = 32$, while a negative $f_\mathrm{NL}\approx-10$ may reduce the arc abundance by $\approx5\,\%$.

\section{A revival problem: Einstein radii and over-concentrations}
Relatively recent observations of strong lensing clusters have reinforced the debate on arc statistics.  Indeed, two additional inconsistencies between expectations in a $\Lambda$CDM cosmology and observations have emerged. First, some galaxy clusters have very extended Einstein rings (i.e. critical lines) which can hardly be reproduced by cluster models in the framework of a $\Lambda$CDM cosmology \citep{2008MNRAS.390.1647B,2004ApJ...607..125T}. For example, studying 12 strongly-lensing galaxy clusters from the Massive Cluster Survey at $z > 0.5$, \cite{2011MNRAS.410.1939Z} found that their Einstein radii are about 1.4 times larger than expected in the standard, $\Lambda$CDM cosmology. 
Second, few clusters, for which high quality strong and weak lensing data became available, have concentrations  which are way too large compared to the expectations \citep{2008ApJ...685L...9B,2009MNRAS.396.1985Z}. These evidences push in the same direction of the ``arc statistics" problem, in the sense that they both suggest that observed galaxy clusters are too strong lenses compared to numerically simulated clusters. 

\subsection{Einstein radii}
\label{sect:einst}

The connection between lensing cross section and Einstein ring size arises from the fact that the former is defined as an area surrounding the caustics on the source plane. Through the lens equation, the caustics are mapped onto the critical curves, thus the length of the critical curves must reflect the size of the lensing cross section. 

\begin{figure}[t]
\begin{center}
\resizebox{12cm}{!}{
\includegraphics{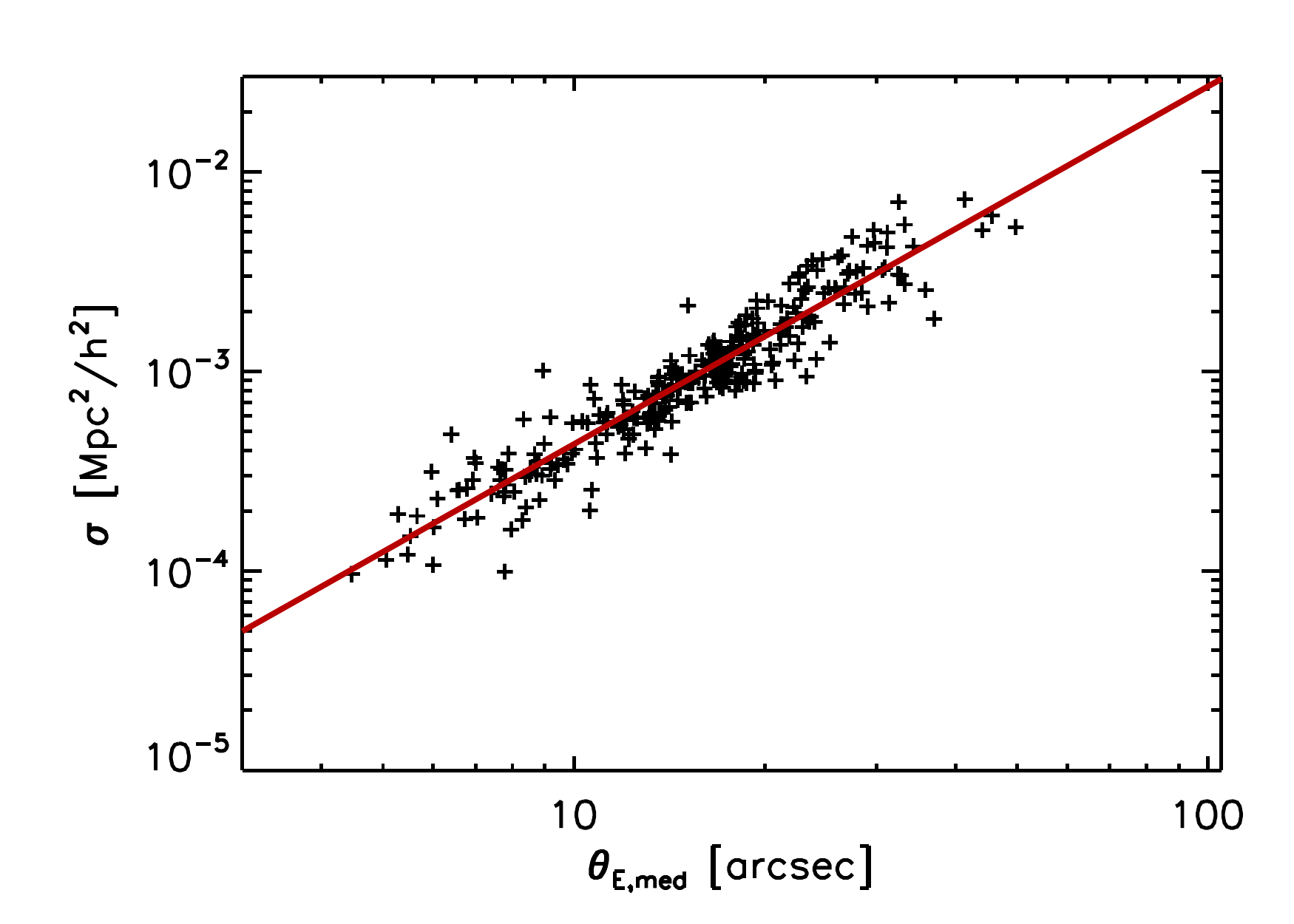}
} 
\end{center}
\caption{Correlation between the cross section for giant arcs ($l/w>7.5$) and the {\em median} Einstein radius, as found by \cite{2011A&A...530A..17M} analyzing a sample of halos with mass $M>5\times10^{14}h^{-1}M_{\odot}$ at $z>0.5$ in the {\sc MareNostrum Universe}.}
\label{fig:thetacs}
\end{figure}

The correlation between these two quantities is shown in Fig.~\ref{fig:thetacs}, taken from \cite{2011A&A...530A..17M}. Each cross corresponds to a halo with mass $M>5\times10^{14}h^{-1}M_{\odot}$ at $z>0.5$ in the {\sc MareNostrum Universe} ($\Lambda$CDM cosmology with $\Omega_0=0.3$, $\Omega_{0,\Lambda}=0.7$ and $\sigma_8=0.9$). The data-points are well fitted by a linear relation in the $\log-\log$ plane, whose equation is
\begin{equation}
\log(\sigma [h^{-2}\mathrm{Mpc}^2])=(1.79\pm 0.04)\log(\theta_{\rm E,med}[\mathrm{arcsec}])-(5.16\pm0.05) \;.
\end{equation}
\cite{2011A&A...530A..17M} propose to quantify the size of the Einstein radius by taking the median distance of the tangential critical points from the cluster center (here indicated as $\theta_{\rm E,med}$). This definition is different from others used in the literature. For example, the {\em equivalent} Einstein radius, $\theta_{\rm E,eqv}$, is defined as the radius enclosing a mean convergence of one \citep{2011MNRAS.410.1939Z,2010MNRAS.404..325R}.  \cite{2011A&A...530A..17M} note that $\sigma$ correlates better with $\theta_{\rm E,med}$ than with $\theta_{\rm E,eqv}$, and explain this difference as due to the fact that $\theta_{\rm E,med}$ better captures the important effect of shear caused by the cluster substructures, whose effect is that of elongating the tangential critical lines along preferred directions, pushing the critical points to distances where $\kappa$ is well below unity \citep[see also][]{BA95.1}. 

\cite{2012A&A...547A..66R} compare $\theta_{\rm E,med}$ to another definition of geometrical nature. Let A denote the area enclosed by the tangential critical curve. Then, the {\em effective} Einstein radius $\theta_{\rm E,eff}$ is defined as 
\begin{equation}
\theta_{\rm E,eff}\equiv \sqrt{\frac{A}{\pi}} \;,
\label{eq:effre}
\end{equation}
i.e. as the radius of the circle having the area $A$. Analyzing halos with analytic, NFW-like mass profiles \cite{2012A&A...547A..66R} find that  $\theta_{\rm E,eff}$ and $\theta_{\rm E,med}$ are tightly correlated, indicating that both definitions could be used to infer the strong lensing cross sections. The best-fit relation between $\sigma$ and $\theta_{\rm E,eff}$ is given by
\begin{equation}
\log(\sigma)=(2.40\pm0.04)\log(\theta_{\rm E,eff})-(5.35\pm0.03) \;.
\end{equation}

Compared to  \cite{2011A&A...530A..17M}, they find a steeper best linear fit between $\log(\sigma)$ and $\log(\theta_{\rm E,med})$:
\begin{equation}
\log(\sigma)=(2.44\pm0.03)\log(\theta_{\rm E,med})-(5.68\pm0.03) \;.
\label{eq:redlich}
\end{equation}
They conclude that the presence of substructure and cluster mergers in the \cite{2011A&A...530A..17M} simulated clusters sample - as opposed to semi-analytic smooth triaxial cluster-halo models - results in shallower slope and higher normalization. Indeed, \cite{2012A&A...547A..66R}  note that during a merger none of the definitions of Einstein radii is able to preserve the tight correlation with lensing cross sections. In particular, the correlation seems to break down completely if {\em median} Einstein radii are used. During the mergers the pairs ($\theta_{\rm E,med}$,$\sigma$) systematically lie below the relation in Eq.~\ref{eq:redlich}, found for isolated clusters. In comparison, {\em effective} Einstein radii result to be less sensitive to cluster dynamics. 

\cite{2012A&A...547A..66R} also show that the slope of the relation between $\log(\sigma)$ and $\log(\theta_{\rm E,med})$ is  sensitive to the inner slope of the cluster density profile, being shallower for halos with steeper density profiles. The effect is of the order of $\sim 5-10\%$ using semi-analytic cluster models. \cite{2012MNRAS.427..533K} compare the correlation between $\sigma$ and $\theta_{\rm E,med}$ for a large sample of halos taken from numerical simulations with different gas physics \citep{2010MNRAS.401.1670F,2011MNRAS.418.2234B}. Clusters from radiative simulations follow a shallower $\log(\sigma)$-$\log(\theta_{\rm E,med})$ relation than clusters with AGN feedback, reflecting the trend expected from the results of \cite{2012A&A...547A..66R}: in absence of efficient energy feedback, gas cooling leads to the formation of steep central density profiles, having a nearly isothermal slope.

\begin{figure}[htbp]
\begin{center}
  \includegraphics[width=0.55\hsize]{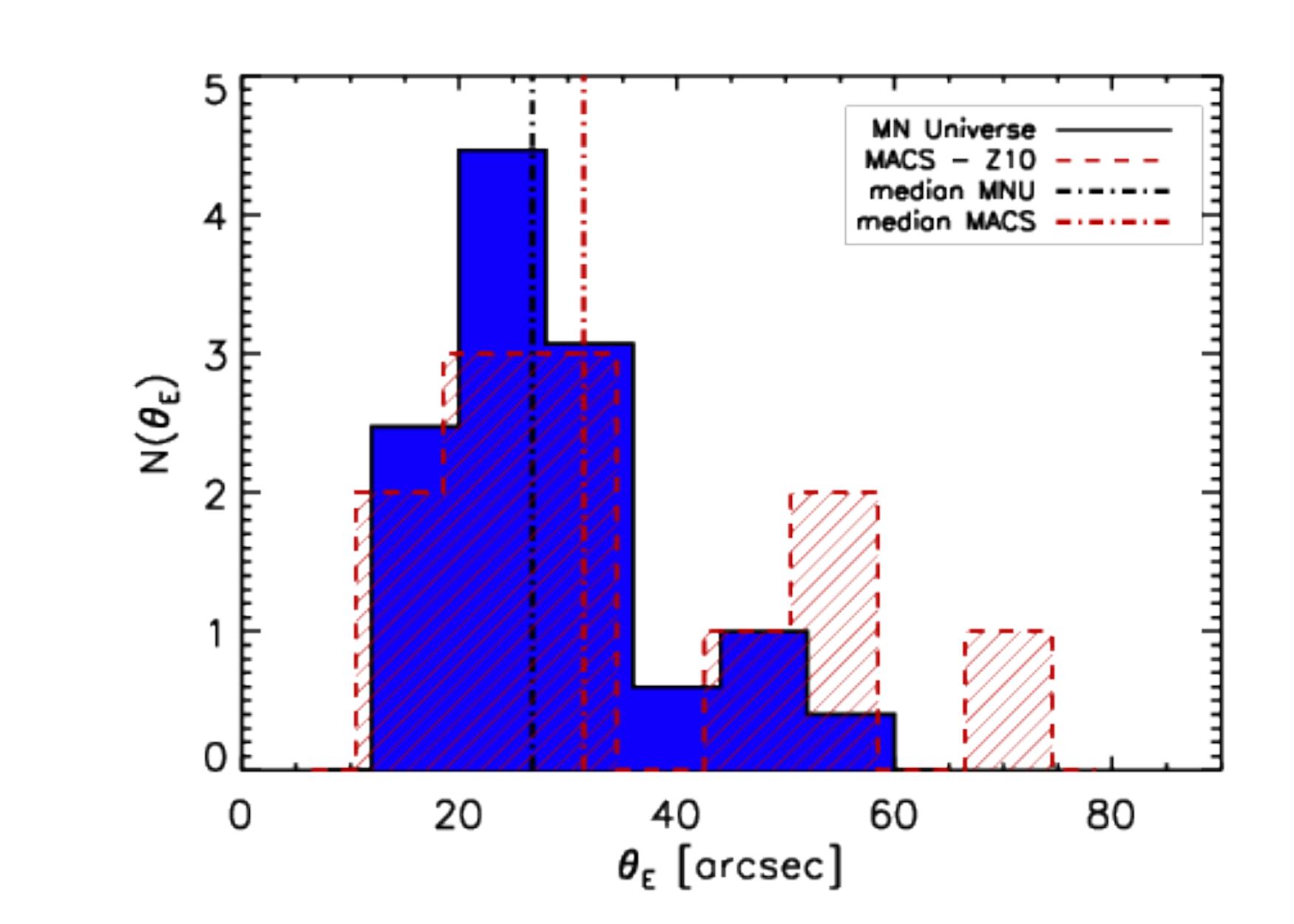}
  \includegraphics[width=0.44\hsize]{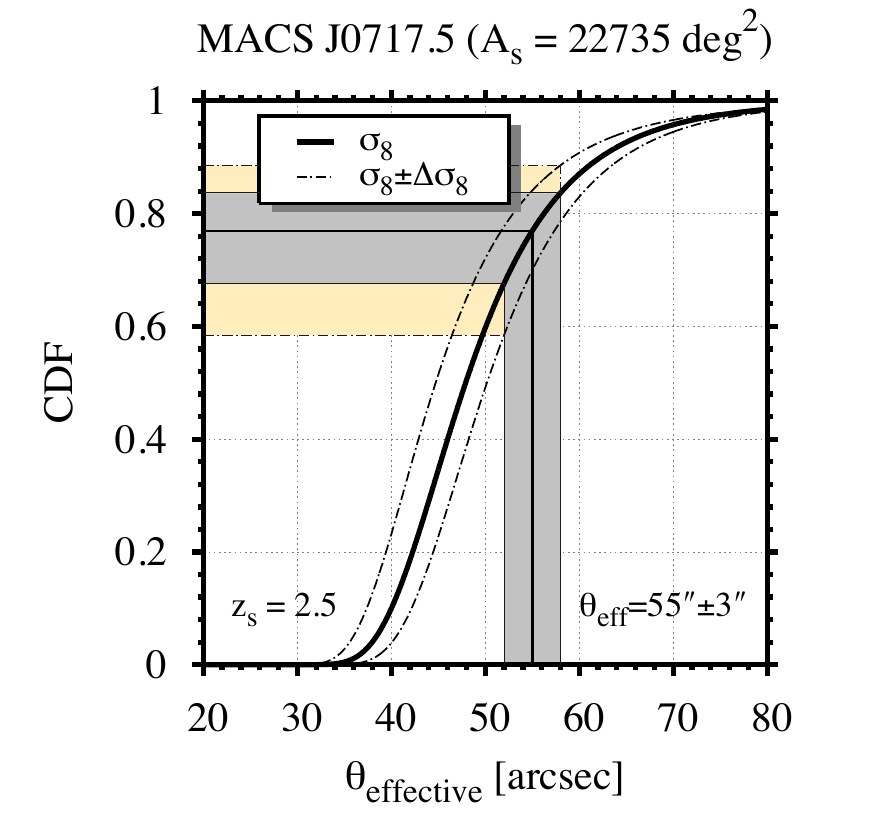}	
\caption{Left panel: The distributions of the Einstein radii in simulated (blue histogram) and observed (red-shaded histogram) MACS samples. The simulated sample was constructed with clusters taken from the {\sc MareNostrum Universe}. The vertical dot-dashed lines indicate the medians of the two distributions. Figure taken from \cite{2011A&A...530A..17M}. Right panel: cumulative distribution function of the largest effective Einstein radius assuming a redshift interval of $0.5<z< 1$ and the nominal MACS survey area. the shaded grey vertical band marks the size of the Einstein radius of MACSJ0717.5+3745 together with its uncertainty ($55''\pm 3''$). The dashed-dotted lines, together with the yellow shaded area, illustrate the impact of the uncertainty in the WMAP7 value of $\sigma_8$ on the cumulative distribution function. Figure from \cite{2012A&A...547A..67W}.}
\label{fig:einstrstat}
\end{center}
\end{figure}

First attempts to compare the observed Einstein radii statistics to simulations were  controversial, though. This is probably due to the limited number of strong lenses available. \cite{2011A&A...530A..17M} compared the statistics of Einstein radii in the {\sc MareNostrum Universe} to those in a sample of 12 MACS clusters at redshift $z>0.5$ \citep{2007ApJ...661L..33E}. They find that the distribution of Einstein radii in the observed sample is characterized by an excess of clusters towards the large values of $\theta_{\rm E}$ (see the left panel in Fig.~\ref{fig:einstrstat}. To be more precise, running a Kolmogorov-Smirnov test, to compare the MACS and the simulated samples, they find a probability of $\sim30\%$ that the two datasets are drawn from the same statistical distribution. The medians of the Einstein radii in the two samples differ by $\sim25\%$. Comparing the MACS clusters to semi-analytical models in the framework of a WMAP7 cosmology, \cite{2011MNRAS.410.1939Z} report a difference of $\sim40\%$ between the observed and the theoretical distributions of Einstein radii. 

As said, the sample investigated by   \cite{2011A&A...530A..17M} consists of only 12 clusters. The tension between simulations and observations arises from three clusters, namely MACSJ0717.5+3745, MACSJ0025.4-1222, and MACSJ2129.4-0741. In particular, MACSJ0717.5+3745 is a very complex system, known to possess the largest Einstein ring on the sky.  Given the small number of systems considered, the question arises whether or not the standard cosmological model can be questioned on the basis of a (nearly) single observed extreme galaxy cluster. In a recent paper, \cite{2012A&A...547A..67W} evaluate the occurrence probability of the Einstein radius of MACSJ0717.5+3745 in a $\Lambda$CDM cosmology using the Extreme Value Statistics (EVS) approach. Very shortly, this statistical approach models the stochastic behavior of the extremes and allows to quantitatively evaluate how frequent unusual observations are. Using triaxial NFW halo models and generating cluster distributions from mass functions calibrated on numerical simulations \citep[e.g.][]{2008ApJ...688..709T},  \cite{2012A&A...547A..67W} determine the expected distribution of Einstein radii, and fit the results with the general extreme value distribution. Although they find that  the distribution of the maximum Einstein radius is particularly sensitive to the precise choice of the halo mass function, lens triaxiality, the inner slope of the halo density profile and the mass-concentration relation, they conclude that MACSJ0717.5+3745 is not in tension with $\Lambda$CDM \citep[see also ][]{2009MNRAS.392..930O}. This is illustrated in the right panel of  Fig.~\ref{fig:einstrstat}: assuming the MACS survey area and adopting the definition of {\em effective} Einstein radius given in Eq.~\ref{eq:effre}, the occurrence probability of an Einstein radius as large as that of MACSJ0717.5+3745 ($55''\pm 3''$) is of (11-42)\%. This range takes into account both  uncertainties on the size of the Einstein radius and on the value of cosmological parameters such as $\sigma_8$. 

\subsection{Over-concentrations}
\begin{figure}[t]
\begin{center}
  \includegraphics[width=1\hsize]{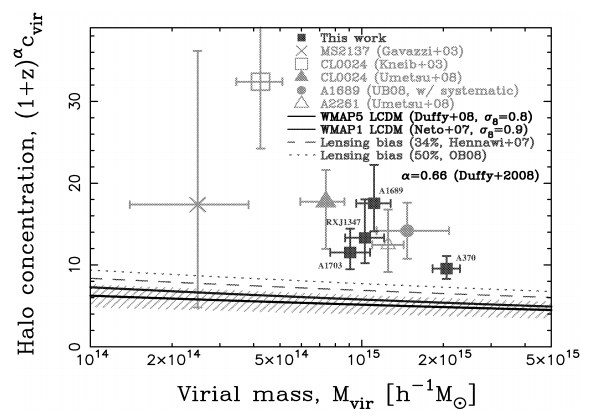}	
\caption{The concentration-mass of few well studied lensing clusters compared to predictions from numerical simulations \citep[e.g.][]{2008MNRAS.390L..64D}. Figure from \cite{2008ApJ...685L...9B}.}
\label{fig:postman}
\end{center}
\end{figure}

The density profiles of galaxy clusters can be well studied by means of several probes. These include observations of strong and weak lensing effects, X-ray emission, and galaxy dynamics (see e.g. the companion reviews by Hoekstra et al., Ettori et al. and Bartelmann et al.). It is a puzzling and potentially highly important problem that combined strong- and weak-lensing analyses of galaxy clusters in many cases find that NFW density profiles well reproduce  the lensing observables, but with concentration parameters that are substantially larger than expected from numerical simulations \citep{BR05.1, 2007MNRAS.379..190C, 2008ApJ...684..177U, 2008ApJ...685L...9B, 2010ApJ...714.1470U,2010MNRAS.403.2077S}. The concentration in this case is defined as the ratio between the virial radius and the scale radius of the profile, which is the radius where the shallower slope in the core turns into the steeper slope farther out. This so-called over-concentration problem is illustrated in Fig.~\ref{fig:postman}, which shows the location in the $c-M$ plane of several massive, well studied galaxy clusters, compared to the predictions from N-body simulations \citep[e.g.][and others]{2008MNRAS.390L..64D}. Some clusters such as A~1689 seem to be extraordinarily concentrated \citep[see however the companion review by][]{2012arXiv1210.3067L}. Whether this is a significant contradiction to the $\Lambda$CDM model remains to be clarified and is currently much debated, in particular because other studies find concentration parameters in the expected range, sometimes in the same clusters
\citep{HA06.1, LI08.1, OK10.1, SE12.1}.   

Due to selection biases and projection effects, strongly lensing clusters should be among the most concentrated clusters, in the sense that concentrations inferred from their projected mass distribution, $c_{\rm 2D}$, should be significantly higher than in three dimensions $c_{\rm 3D}$. \cite{2010A&A...519A..90M} show that this concentration bias is a heavily conditioned by the lens redshifts and strong lensing cross sections. For example, at intermediate redshifts the ratios between the 2D  and the corresponding 3D concentrations of critical clusters (i.e. clusters capable of producing multiple images) are $\sim 1.2$, but they become higher than 1.4 at low and at high redshifts. For the strongest lenses with cross sections for giant arcs $\sigma>10^{-3}h^{-2}$Mpc$^{2}$, the concentration bias $c_{\rm 2D}/c_{3D}$ can be of the order of $50-70\%$ even for clusters with mass $M>7\times 10^{14}h^{-1}M_\odot$ at any redshift. These results agree well with those of \cite{2009MNRAS.392..930O}, who use semi-analytic models of triaxial halos to estimate that the projected mass distributions of strong lensing clusters have $\sim 40-60\%$ higher concentrations than typical clusters with similar redshifts and masses (see also Sereno et al. 2010). Other studies suggest that the  extreme concentrations  of some galaxy clusters derived from the lensing analysis can be explained by means of cluster elongation along the line of sight \citep[see the companion review by][and the references therein]{2012arXiv1210.3067L}.

Whether cluster triaxiality is sufficient to fully explain the observed lensing concentrations is yet unclear. Indeed, the bias depends strongly on the cluster selection method. For example, comparing the halos from the {\sc MareNostrum Universe} to MACS clusters, which constitute a sample of X-ray selected rather than lensing selected clusters, \cite{2011A&A...530A..17M} estimate that the median concentration bias of the MACS sample is only $\sim 11\%$. For $\sim 20\%$ of the sample the concentration bias is expected to be $>40\%$. 

Among the foremost goals of the ongoing Hubble Multi-Cycle Treasury programme CLASH \citep{2012ApJS..199...25P} is the measurement of accurate concentration parameters in sample of 25 massive X-ray bright galaxy clusters. This program has been designed to shed light on the over-concentration issue and is already producing interesting results on this subject (see discussion in the following section).

\section{On-going gravitational arc surveys}

\subsection{CLASH}
The {\em Cluster Lensing and Supernova survey with Hubble}\footnote{http://www.stsci.edu/~postman/CLASH/Home.html} (CLASH) Multi-Cycle Treasury program \citep{2012ApJS..199...25P} has been awarded a total of 524 orbits of time on HST to observe 25 clusters over a 3 years period, and is currently delivering images of each cluster in 16 HST bands (from the UV to the IR wavelengths), making use of the refurbished Advanced Camera for Surveys and of the new Wide Field Camera 3. The HST observations are complemented with wide-field observations from several ground-based facilities, including the Subaru telescope, enabling the weak-lensing analysis of the same clusters. An ESO large-program (PI: Piero Rosati) is currently ongoing to provide spectroscopic follow-up of both cluster members, gravitational arcs, and multiple images in the CLASH clusters with the VLT. All these observations are complemented with X-ray (Chandra and XMM), IR (Spitzer), and SZE (Bolocam, Mustang) data.

This is the first sample of clusters for which homogeneous, high resolution, and deep imaging becomes available. Each cluster reveals several multiple image systems, arcs and arclets , which are used to  constrain mass models and to measure several properties of the clusters cores, including their Einstein rings and their strong lensing cross sections \citep{2012ApJ...749...97Z,2012ApJ...757...22C,2012ApJ...755...56U}.  The combination of weak- and strong-lensing allows robust measurements of the mass profiles and of the cluster concentrations. The CLASH sample is thus ideal  for arc-statistics studies in many respects. 20 out of the 25 clusters in the sample were selected on the basis of their X-ray temperature ($T_X>7$ key) and X-ray morphology. The remaining 5 clusters were selected because of their extremely large Einstein radii. 

\begin{figure}[htbp]
\begin{center}
  \includegraphics[width=0.42\hsize]{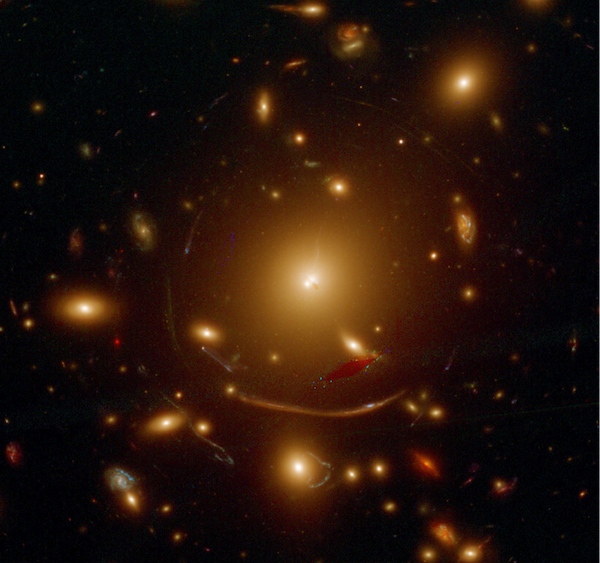}
  \includegraphics[width=0.57\hsize]{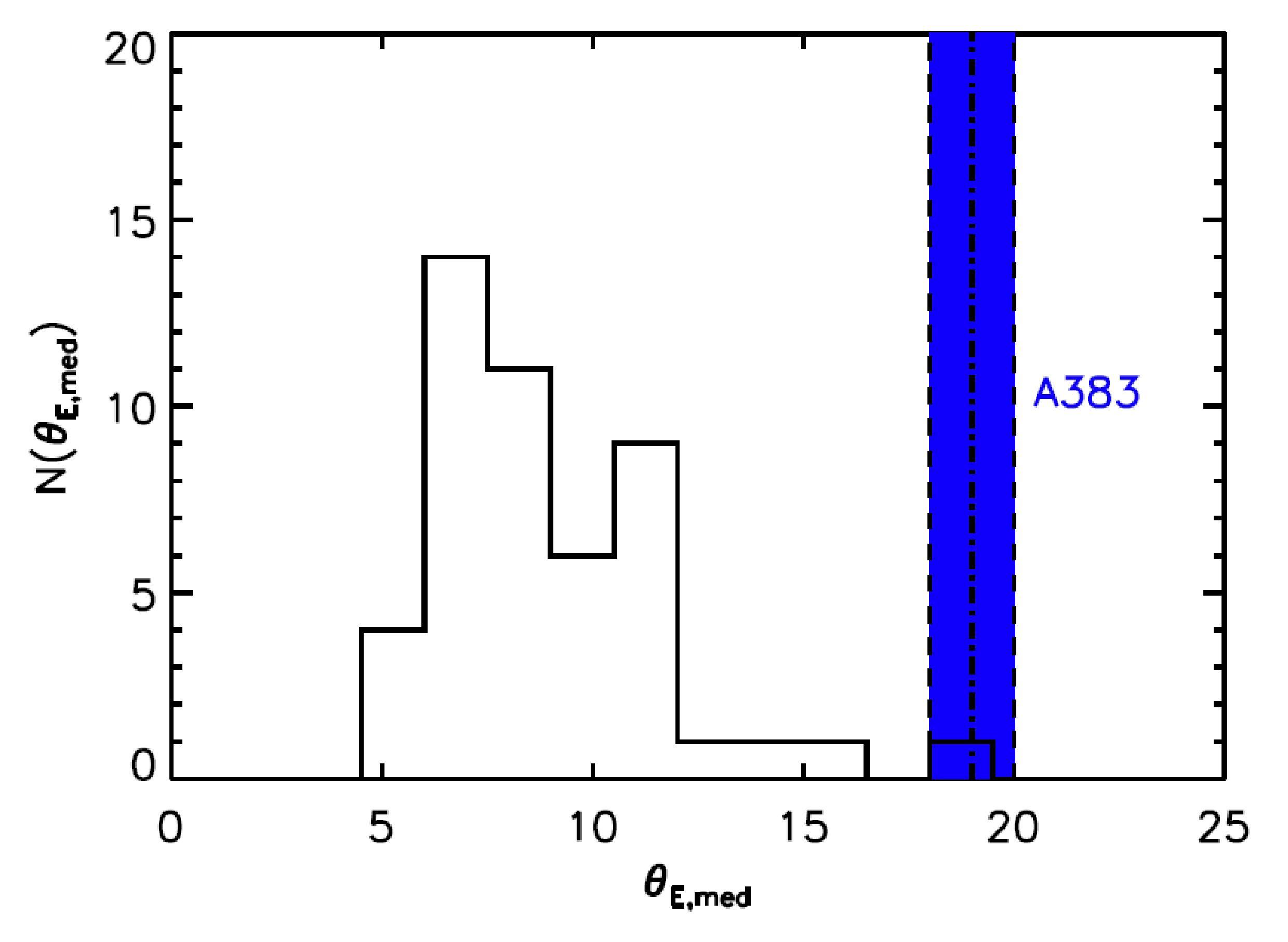}	
\caption{Left panel: CLASH observation of the galaxy cluster Abell 383. Right panel: the black histogram shows the distribution of Einstein radius sizes for clusters with mass similar to that of A383, extracted from the $\Lambda$CDM cosmological simulation {\tt MareNostrum Universe} \citep[see e.g.][]{2010A&A...519A..90M}. The size of the Einstein radius of A383 and its error are given by the blue vertical stripe.}
\label{fig:a383}
\end{center}
\end{figure}

The first cluster observed by the CLASH collaboration was Abell 383 \citep{2011ApJ...742..117Z}, see left panel of Fig.~\ref{fig:a383}. The strong lensing model constructed on the basis of 27 multiple images of 9 different sources at redshifts $1.01 < z< 6.03$ has an Einstein radius of $\sim 19"$  and a cross-section for giant arcs of $2.26\times10^{-3}h^{-2}$ Mpc$^2$ (both for $z_{rm s}=2$). These values are apparently not overwhelmingly large \citep[see e.g. the values quoted in][]{2011A&A...530A..17M}. However, the mass of this cluster at $z=0.187$, estimated by combining weak and strong lensing data, is relatively small ($M_{vir}\sim 5.5\times10^{14}h^{-1}M_{\odot}$). Halos of similar mass and redshift in the hydrodynamical simulation {\sc MareNostrum Universe} ($\Lambda$CDM cosmology with $\Omega_{\rm m}=0.3$, $\Omega_{\Lambda}=0.7$ and $\sigma_8=0.9$,\citealt[see ][]{2010A&A...519A..90M}) have significantly smaller Einstein radii and cross sections, as shown in the right panel of Fig.~\ref{fig:a383}. The lensing data also suggest that the concentration of Abell 383 is $c_{vir}=8.8$, which is a factor of $\sim 1.75$ higher than the median concentration of numerically simulated halos at this mass scale. Accounting for the expected lensing bias ($\sim 35\%$) still does not remove the tension with $\Lambda$CDM completely. However, it is worth noting that the scatter in concentration at fixed mass if found to be large in numerical simulations, where concentrations typically have log-normal distributions with dispersion $\sigma\sim0.3$. As shown in \cite{2012arXiv1210.3067L}, the estimated concentration of this cluster can be reduced to $\sim 5$ by assuming that the cluster is elongated along the line of sight.

 The combined weak and strong lensing analysis of the cluster Abell 2261 ($z=0.22$) suggests that the concentration and mass of this cluster are $c_{vir}\sim6.2$ and $M_{vir} \sim3.1\times10^{15}h^{-1}M_\odot$, respectively \citep{2012ApJ...757...22C}. This is $\sim 40-50\%$ above the expected $c-M$ relation for $\Lambda$CDM \citep{2008MNRAS.390L..64D,2009ApJ...707..354Z}. Previous analyses based on weak-lensing data only found concentrations in the range $c_{vir}\sim6-10$ for this cluster \citep{2009ApJ...694.1643U,OK10.1} Using {\em Chandra} X-ray data \cite{2012ApJ...757...22C} also estimate the cluster mass under the assumption of axial symmetry and hydrostatic equilibrium of the intra-cluster medium. They find that the X-ray mass is lower by $\sim35\%$ than the lensing mass at $R_{2500}$. By assuming that the mismatch between mass estimates can be attributed to cluster triaxiality, they compute the expected elongation of the cluster mass distribution along the line of sight. Accounting for the elongation necessary to explain the mass mismatch, they show that the concentration and mass of Abell 2261 can be brought in full agreement with the theoretical predictions in the framework of the $\Lambda$CDM cosmology.    

Again using the CLASH HST and Subaru data, \cite{2012ApJ...755...56U} derive the mass profile of the cluster MACS J1206.2-0847 ($z=0.439$) combining weak-lensing distortion, magnification, and strong lensing constraints. They find that the cluster is well described by an NFW density profile with $M_{vir}\sim 1.1h^{-1}M_\odot$ and $c_{vir}\sim 7$. The expected concentration \citep[see again][]{2008MNRAS.390L..64D,2009ApJ...707..354Z}at this mass scale is $4-5$. Differently from Abell 2261, the lensing and the X-ray mass estimates of this cluster agree well.   

The analysis of other CLASH clusters is still in progress.

\subsection{The SOAR Gravitational Arc Survey}
The SOAR Gravitational Arc Survey \citep{2012arXiv1210.4136F} is a survey conducted with the Southern Astrophysical Research Telescope (SOAR). During the period from mid 2008 to the end of 2010, it has delivered observations of 47 galaxy clusters optically selected in the Sloan Digital Sky Survey (SDSS) Stripe 82 (40 clusters) and in an additional field at higher RA in an equatorial region covered by SDSS single pass imaging \citep{2012arXiv1210.4136F}. Additional 4 clusters are selected from the catalog of ROSAT X-ray sources, just because they had good observability with the SOAR telescope during the allocated semester of observations/ The clusters have photometric redshifts in two narrow intervals centered at $z=0.27$ (22 clusters) and $z=0.55$ (25 clusters), and are imaged in three bands down to magnitude limits of $g\sim 23$, $r\sim22.5$, and $i\sim 22$ for $S/N=3$. One of the goals of the survey is verify with a better statistics the results of \cite{GL03.1} and \cite{ZA03.1}, showing that the incidence of giant arcs is higher at higher redshifts. Gravitational arcs are searched by visually inspecting the stacked $g+r+i$ coadded images. The median seeing for all images is $0.82"$, $0.74"$, and $0.69"$ in the $g$, $r$, and $i$ bands, respectively.

\cite{2012arXiv1210.4136F} report that 6 cluster fields clearly contain gravitational arcs, while they find arc candidates in additional two clusters. The total number of detected arcs is 16, with 4 of them being classified as {\em giant}, having $l/w>7$. Two of the clusters containing arcs belong to the sample of X-ray selected systems. Of the remaining 6 optically selected clusters forming arcs, four of them are in the high-$z$ bin, and 2 are in the low-$z$ bin, which is qualitatively in agreement with the results of \cite{GL03.1} and \cite{ZA03.1}. The higher incidence of arcs among the X-ray selected clusters than among the optically selected ones is consistent with the results of \cite{2010MNRAS.406.1318H}.

%The sample contains 6 clusters with   
%- Clusters are divided in two redshift bins: 0.2 < z < 0.35 and 0.5 < z < 0.6
%- Clusters have photo-zs
%- SOAR telescope, depth of observations:   g',r',i'<23, 22.5, 22 (S/N=3)
%- clusters are selected by richness but the richness or mass cut was never mentioned
%- found 6 clusters with arcs + 2 clusters with  candidates. 16 arc candidates in total. 4 arc candidates have l/w>7. Not all these clusters are selected by richness!!!
%- two of the lenses are in the low-z bin, 4 in the high-z bin (richness selected)
%- efficiency measured: $4/25 = 16\pm8\%$ in the high-z bin; $10\%\pm6\%$ for the low-z bin;
%- authors claim this is consistent with Gladders et al. 2004 and Hennawi et al. 2008, but there is no discussion about how the samples can be compared (depth, filters, seeing, etc...)

\subsection{The Sloan Bright Arcs Survey}
The Sloan Bright Arcs Survey (SBAS) is a survey focused on the discovery of strong
gravitational lensing systems in the SDSS imaging data and on subsequent analysis of these systems \citep[e.g.][]{2007ApJ...662L..51A,2009ApJ...707..686D,2009ApJ...696L..61K}. Lenses are identified by looking for blue objects ($g-r<1$ and $r-i<1$) around a catalog of $29,000$ Brightest Cluster Galaxies (a parallel investigation targets 221,000 Luminous Red Galaxies to search galaxy-scale strong lensing systems). To date, 19 strongly lensed galaxies have been discovered and spectroscopically confirmed in the data.

An interesting result form this survey is described in paper by \cite{2012ApJ...761....1W}, who  took follow-up at the WIYN telescope of 10 cluster-scale systems exhibiting arc-like features. For each cluster in this strong-lensing selected sample, they measure the richness (i.e. the number of cluster members within $1\;h^{-1}$Mpc of the BCG) by means of the maxBCG method \citep{0004-637X-633-1-122}. The richness is used as a proxy to determine the cluster mass  and the velocity dispersion, using the empirical relations found by  \cite{2007arXiv0709.1159J} and by \cite{2007ApJ...669..905B}. They estimate the Einstein radii by fitting circles to the visible arcs and measuring the radii of such circles. The mass enclosed with the Einstein radius is estimated using Eq.~\ref{eq:13}. They further estimate the velocity dispersion within the Einstein radius by approximating the clusters to singular isothermal spheres. In this case,
\begin{equation}
 \sigma_v=\sqrt{\frac{\theta_E c^2D_{s}}{4\pi D_{ls}}} \;.
\end{equation}
Comparing these velocity dispersion measurements to those obtained from the richness (which refer to $R_{200}$, i.e. to much larger radii compared to the Einstein radii), they find that, on average, the estimates agree with each other for most of the clusters, suggesting that these systems are largely isothermal. However, they claim to find some segregation between low-mass and high-mass systems. For the former, they find velocity dispersions decreasing from the inner to the outer radii, indicating that these clusters have a large fraction of their mass concentrated in the clusters cores. On the contrary, high-mass clusters tend to exhibit the opposite trend, indicating that much of the mass is found at larger distance from the BCG and suggesting low values of the concentration. 

\begin{figure}[t]
\begin{center}
  \includegraphics[width=1.0\hsize]{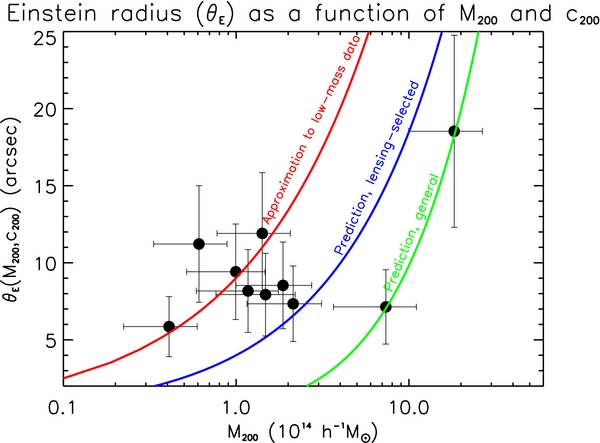}
\caption{From Wiesner et al. (2012): Plot of Einstein radius vs. $M_{200}$ for the ten clusters with giant arcs in the SBAS survey. See text for more details.}
\label{fig:sbas}
\end{center}
\end{figure}

They also use the cluster masses to derive the expected sizes of the Einstein radii under the assumption that clusters have an NFW density profile and that the $c-M$ relation of \cite{2008MNRAS.390L..64D} holds. The results are shown in Fig.~\ref{fig:sbas}. They find that the Einstein radii of the SBAS clusters are significantly larger than expected from NFW halos with equivalent masses (data-points vs green solid line). The comparison is done by rescaling the observed Einstein radii to a common set of source and lens redshifts. They note that the clusters that deviate mostly from the expectations are those with the lowest masses and they argue that using the $c-M$ relation corrected for the lensing bias by \cite{2009ApJ...699.1038O} (blue solid line) still does not bring the data in agreement with the predictions. The red solid line in the figure denotes the best-fit relation to the four clusters with the lowest masses in the sample. Using the measured values of the Einstein radii and masses of these last clusters, the average concentration is $c_{200}\sim 11$, i.e. almost twice the value expected from numerical simulations in the framework of the $\Lambda$CDM cosmology.      
\cite{2012ApJ...761....1W} note, however, that the richness measurements based on the follow-up imaging with the WIYN telescope are lower than those obtained with the SDSS data, and they attribute this inconsistency to the large photometric errors affecting the shallow SDSS observations. This might have caused the mis-identification of several field-galaxies as cluster members. If they scale the richness values to match the SDSS values in order to use the mass-richness relation of \cite{2007arXiv0709.1159J}, the mass estimates are higher by a factor $\sim 2$. Shifting the data-points by this amount along the $M_{200}$-axis in Fig.~\ref{fig:sbas} brings the majority of the SBAS clusters to match the predictions for the lensing-biased $c-M$ relation. We add  that part of the  inconsistency between observations and theoretical expectations may be originated here by the assumption of axial symmetry of the lenses  made by the authors of this paper. Indeed, if the radial symmetry of the lens is perturbed, the tangential critical line becomes elongated. As noted by \cite{ME05.2}, the longest arcs tend to form along those portions of the critical lines that are at the largest distance from the cluster center \citep[see also][]{DA03.1}. Circles traced through these arcs do not enclose a mean convergence of unity (see Sect.~\ref{sect:axsym} and \citealt{BA95.2}) and, thus, the Einstein radii measured in this way are over-estimated by an amount that depends on the ellipticity of the lenses. Additionally, the richness-mass relation is affected by a large scatter, which may also impact of the analysis described above. 

%Paper by Wiesner et al. (2012):
%- 10 clusters from the SDSS where arcs have been identified via color selection (blue features around bright cluster members).
%- follow-up with the WIYN telescope
%- measure richness, mass and velocity dispersion of  clusters
%- do very crude measurement of the Einstein radius assuming spherical symmetry.  
%- use theoretical c-M relations to estimate c from M and then the Einstein radius. They compare the resulting theta_e to their observations 
%- find that low-mass systems are over-concentrated, while other clusters are not.

Another search for giant gravitational arcs in the SDSS is the Sloan Giant Arc Survey (SGAS) of which an overview is given by \cite{2008AJ....135..664H}. The  targets are clusters selected  using the red cluster sequence technique \citep[RCS selection algorithm,][]{2000AJ....120.2148G}, which span the redshift range $0.1 \leq z \leq 0.6$. Given that the area covered by the SDSS DR7 is 8000 sq. degrees, the survey subtends a comoving volume of $\sim 2$ Gpc$^3$. The initial phase of the survey, for which results results were reported by  \cite{2008AJ....135..664H}, carried follow-up imaging observations at the WIYN and at the UH88 telescopes, giving higher priority to the richest clusters. The follow-up observations reach limiting surface brightnesses between $\mu_g=25.7$ and $\mu_g=26.9$ mag arcsec$^-2$. The seeing conditions are rather dishomogeneous (the seeing varies between 0.5" and 2").  The initial analysis shown in \cite{2008AJ....135..664H} refers to 240 clusters  (195 imaged from WIYN and 45 from UH88). After visual inspection of the cluster images done by multiple examiners, who ranked the lensing features with a scale of grades, the authors report that 22 clusters are definite  lenses (of which 6 were previously known), 14 are likely lenses (2 previously known) and 9 are less secure systems. Most of the lens systems are found with the best seeing conditions (median seeing for the whole sample is 0.94", while the median seeing for the lenses is 0.74"). In subsequent phases of the SGAS survey, lens candidates were identified based on a visual inspection of SDSS imaging of RCS-selected clusters. Each cluster was idependently inspected by four examiners and assigned a numerical score based on the likelihood of being a genuine lens. Imaging follow-up to confirm or refute the lensing interpretation of highly-ranked candidates was done using the 2.56m Nordic Optical Telescope. 

Using 28 lensing features detected in the SGAS, \cite{2011ApJ...727L..26B} studied the redshift distribution of the sources producing giant arcs. Precise spectroscopic redshifts for these arcs were obtained with the Gemini/GMOS-North \citep[see also][]{2011ApJS..193....8B}. They find a median redshift of $z=1.821$, with nearly two thirds of the arcs arising from sources at $z \gtrsim 1.4$, indicating that sources giving rise to giant arcs are typically at high redshifts. They use these results to construct a model that attempts to predict the redshift distribution of giant arcs as a function of the limiting intrinsic brightness of the background sources. The model involves 1) the scaling relation for the arc cross-sections. $\sigma_{l/w}$, derived by \cite{2010A&A...519A..91F} by analyzing numerical simulations, 2) the assumption of a universal matter density profile slope ($\alpha\sim-1.5$), necessary to model the variation of the cross actions with the source redshift,  and 3) the intrinsic redshift distribution, $dn/dz_s$, of the background sources as given by the photometric redshift catalog of the COSMOS field \citep{2009ApJ...690.1236I}. The resulting formula describing the arc redshift distribution is
\begin{equation}
\label{eq:bayliss}
\frac{dp_{arc}}{dz_s}=\frac{\sigma_{l/w}\frac{dn}{dz_s}}{\int dz_s\sigma_{l/w}\frac{dn}{dz_s}} \;.
\end{equation}
\begin{figure}[t]
\begin{center}
  \includegraphics[width=1.0\hsize]{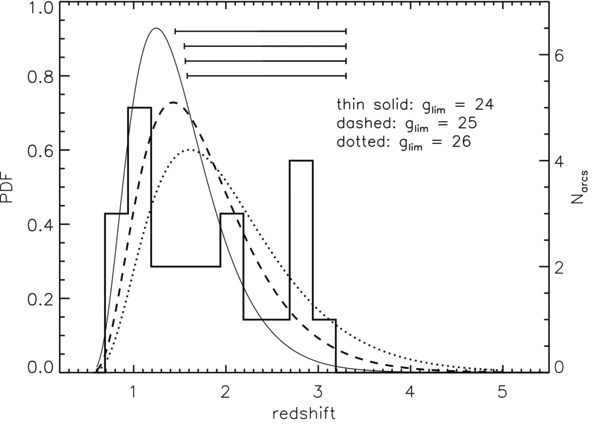}
\caption{From Bayliss et al. (2011): The redshift distribution of 28 giant arcs in the SGAS (histogram). The three solid, dashed, and dotted lines show the predicted redshift distributions, based on the model in Eq.~\ref{eq:bayliss}, for limiting intrinsic magnitudes of the background sources $g=24,25$ and $26$, respectively.}
\label{fig:sgas}
\end{center}
\end{figure}

Recently, \cite{2012ApJ...744..156B} extended the previous work by analyzing a larger sample of arcs (105) from the SGAS and from the Second Red Sequence Cluster Survey (RCS2). The redshift range of the lenses in this enlarged sample is ampler ($0.2<z<1.2$, with a median lens redshift $z_{l}=0.58$). Arcs are subdivided into redshift bins on the basis of their colors (from $u,g,r,z$-band photometry). The results are consistent with the previous study and show that arcs with $g\lesssim 24$ have a median redshift of $z_s \sim 2$. They claim that adopting this median redshift and using the scaling of the optical depth for giant arcs, $\tau_{l/w}$, as given in \cite{WA03.1} would solve the arc statistics problem. However, \cite{LI05.1} and \cite{FE05.1} show that the scaling of $\tau_{l/w}$ and $\sigma_{l/w}$ with source redshift is different from what assumed by \cite{WA03.1}, both in terms of amplitude and of derivative (see discussion in Sect.~\ref{sect:srednorm}).

Blanchard et al.\ (2013, in prep) used SDSS spectroscopy of BCGs in 89 cluster lenses found in the SGAS survey to search for a correlation between spectral diagnostics of star formation/baryonic cooling in the cluster centres and strong lensing efficiency. It was found that the star formation diagnostics in strong lensing clusters showed no statistically significant deviations from the total cluster population. These results argue against the ability of baryonic cooling to strongly modify the strong lensing cross sections of clusters, agreeing well with the results from recent simulations \citep{2012MNRAS.427..533K}. 

  %Papers by Hennawi et al. 2008, Bayliss et al. 2011, Bayliss 2012:
%- H08 introductory paper
%- select clusters with RCS algorithm (red-sequence, over density, magnitude, color) in 8000 sq. deg. area (SDSS DR5).
%- photos for clusters form the red-sequence fitting
%- follow-up with WiYN and UH88 (5 bands -> better photo-z)
%- 240 clusters were followed up. They found 22 definite strong lensing clusters, 14 likely lenses and 9 possible lensing clusters. About ~30 newly discovered lenses.
%- arcs seem to be better found in images with lower seeing (0.74 vs 0.95)
%- depth of the observations is given in surface brightness (between 25.7 and 26.5 for WIYN in the g-band and 26.9 in the V-band for UH88. 
%Bayliss et al (2011) and Bayliss 2012 have used the SGAS data to measure the photometric-redshift distribution of 105 giant arcs.  They find that the median source redshift is $zs=2\pm0.1$, which agrees remarkably well with the redshift distribution measured from spectroscopic observations of a much smaller sample (N = 28) of giant arcs that were identified in much shallower imaging data.
%They claim that adopting this median redshift and using the scaling of tau as given in Wambsganss et al. 2004 would solve the arc statistics problem. However Li et al. (2005) and Fedeli et al (2006) show that the scaling of tau and sigma with source redshift is different from what assumed by Wambsganss (amplitude and derivative).

\subsection{SZ-selected clusters: arcs in clusters detected with the ACT}
The recent developments in the millimiter and sub-millimeter  astronomy, thanks to the advent of instruments like the Atacama Cosmology Telescope (ACT, \citealt{2007ApOpt..46.3444F}), the South Pole Telescope (SPT, \citealt{2011PASP..123..568C}), and Planck \citep{2011A&A...536A...1P}, have enabled in the last years several searches for galaxy clusters through the Sunyaev-Zel'dovich Effect (SZE, \citealt{SU72.1}). This selection method has several advantages compared to optical or X-ray based selections. Indeed, it is relatively independent on redshift (the SZE brightness is redshift independent, but instruments have their own resolution limit), it is less prone to contaminations by foreground or background structures, and, according to simulations, the SZE flux can be related to the cluster mass via relatively clean scaling relations.

\begin{figure}[t]
\begin{center}
  \includegraphics[width=1.0\hsize]{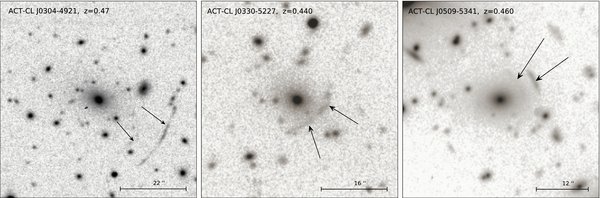}
\caption{From Menanteau et al. (2010): Close-up monochrome images of examples of strong-lensing arcs in three clusters in the ACT sample.}
\label{fig:act}
\end{center}
\end{figure}

Along this line, the ACT and the SPT experiments have so far delivered detections of several cluster candidates, after surveying hundreds of squared degrees of the sky. \cite{2010ApJ...723.1523M} presented a sample of 23 optically confirmed SZE clusters, resulting from the optical follow-up  with the NTT and with the SOAR telescopes of the detections in the 148 GHz data from the ACT 2008 southern survey. It tuned out that 6 out of the 23 clusters show strong lensing arcs (see Fig.~\ref{fig:act}). Of the six clusters, three were already known to be strong lenses, while the others are newly discovered.  

The idea of performing arc statistics studies using SZ-selected clusters is long-standing \citep[see e.g.][]{2001astro.ph..9250M}. Now, it is becoming feasible thanks to the above mentioned experiments. In particular, the {\em Planck} satellite is completing its full-sky survey, and has already delivered a catalog of $\sim 190$ SZE selected clusters. The SPT and the ACT observations have allowed the identification of 240 clusters and 68 clusters, respectively \citep{2013ApJ...763..127R,2013arXiv1301.0816H}.

%To be shifted in the section on selection effects: Meneanteau et al. 2010. Atacama cosmology telescope. Sample of SZ selected clusters (455 sq deg sub map from the ACT 2008 observing season, S/N>3.5 in optimally filtered maps). In six of these systems there is evidence for strong lensing in follow-up optical observations with the NTT and with the SOAR telescopes. Could be a very nice sample of arcs to be used for arc statistics. 

\subsection{Arcs in the HST archive}
The Hubble Space Telescope archive contains plenty of high-resolution images of galaxy cluster fields. These observations represent a treasury to be explored looking for gravitational arcs. Dealing with such an heterogeneous sample of observations makes difficult to predict reliably the  expected number of detectable arcs as a function of cosmological parameters, which depends on the lens selection function and on several observational issues (e.g. the surface brightness limit, etc.). Nevertheless, several properties of the lenses can be unveiled by investigating the statistical properties of the lensed features. 

For example, \cite{SA05.1} did a systematic search for gravitational arcs in galaxy clusters contained in the Hubble Space Telescope Wide Field and Planetary Camera 2 data archive. They restricted the analysis to potential lenses with known redshifts in the range $0.1<z<0.8$ and available observations in one or more of the following bandbs: F450W, F555W, F606W, F675W, F702W, and F814W. Arc-like features were searched by visual inspection, classifying them into radial and tangential arcs. Given that radial arcs are generally located close to the cluster center \cite{SA05.1} examined the images after subtracting the central brightest galaxies. In total, 128 clusters were examined. They found  12 candidate radial arcs and 104 tangential arcs having a length-to-width ratio exceeding 7. 

As suggested by several authors \citep[e.g.]{2001ApJ...559..544M,OG01.1,WY01.1} the core structure of dark-matter halos can be probed with tangential and radial arcs. The idea is based on the fact that the tangential and radial magnifications probe the enclosed mass and the slope of the mass profiles, respectively (see Sect.~\ref{sect:axsym}). In particular, nearly isothermal profiles have radial magnification $\mu_r\sim1$, implying that no radial arcs can be produced with such steep density profiles. \cite{OG02.1} showed that the ratio of radial-to-tangential arc numbers is particularly robust against the various systematics that affect the cross section for lensing. Following this idea, \cite{SA05.1} compared the relative  abundances of radial to tangential arcs in the HST WFC2 archive to expectations for toy models based on the combination of a main dark-matter halo with generalized NFW profile,
\begin{equation}
\rho(r)=\frac{\rho_s}{(r/r_s)^\beta(1+r/r_s)^{3-\beta}}\;,
\end{equation}
and a central luminous component (the BCG), described by the \cite{1990ApJ...356..359H} profile
\begin{equation}
\rho_H(r)=\frac{M_L r_H}{2\pi(r_H+r)^3} \;,
\label{eq:gnfw}
\end{equation}
where $M_L$ is the total luminous mass and $r_H$ is related to the BCG effective radius: $R_e=1.8153r_H$. They found that the arc statistics of this HST survey are consistent with a range of density profiles with $\beta \lesssim 1.6$. The precise value of $\beta$ depends strongly on the assumptions made on the BCG mass. 

\section{Future prospects}
As outlined in the previous sections,  the persistence of the {\em arc statistics problem} is still questioned both by theoreticians and by  observers. The main reasons for such uncertain picture are 1) the lack of a proper interface between theoretical calculations and observations; 2) the lack of homogeneous cluster surveys, targeting systems with well defined selection functions. To bring arc statistics to the level of a competitive cosmological probe, these two issues need to be addressed. Fortunately, big improvements are possible on both aspects, given our better understanding of strong lensing clusters, which will help us to better model the lenses, and to the recent design of several observational campaigns. In this Section, we will discuss how the status of research in arc statistics is likely to change in the next future. 

\subsection{Improving the interface between theory and observations}
First of all several methodological aspects need to be improved in order to make arc statistics a competitive cosmological tool. For example, are the methods used in simulations to measure the arc properties easily exportable to observations? Answering to this question is rather crucial, because the correct comparison between theory and observations requires that the arc properties are measured consistently.  

The properties of arcs in real observations are obviously more complicated to be measured than in simulations. In particular, a big complication is arising from the fact that arcs form in crowded regions of clusters. The light from cluster members can influence the detectability of the lensed images or of part of them, thus affecting the measurements of the arc widths and lengths. Additionally, the measured length and width depend on the depth of the observation because of the source brightness profile, which typically declines with the distance from the source center, and the background noise. Arc properties are also affected by the instrument PSF and, in case of observations from the ground, by atmospheric blurring. Several of these issues are discussed in \cite{2008A&A...482..403M}.  

Let assume that we can use some software to identify the pixels belonging to an arc above a certain value of the background r.m.s. Measuring the arc length  can  be done in the same way as in ray-tracing simulations, discussed in Sect.~\ref{sect:meth}). One can identify the brightest arc pixel as point (1) (see Fig.~\ref{figure:arcfit}) and then select the point (2) and (3) as explained above. Then, the arc length can be estimated by tracing a circular segment through points (1), (2), and (3).

Measuring the arc widths is more tricky because: a) real arcs are originated by sources that are not simple ellipses. They are typically spiral or irregular galaxies at high redshift, which are characterized non-uniform brightness profiles, multiple bright knots, and asymmetric shapes. Arcs originate from mergers of multiple images of the same source. Thus, the width is not constant along the arc.; b) most arcs have low surface brightnesses and they barely emerge from the background noise. Thus, their edges are very irregular and the arc perimeter is a noisy quantity.

\begin{figure}[t]
\begin{center}
\resizebox{12cm}{!}{\includegraphics[width=0.47\hsize]{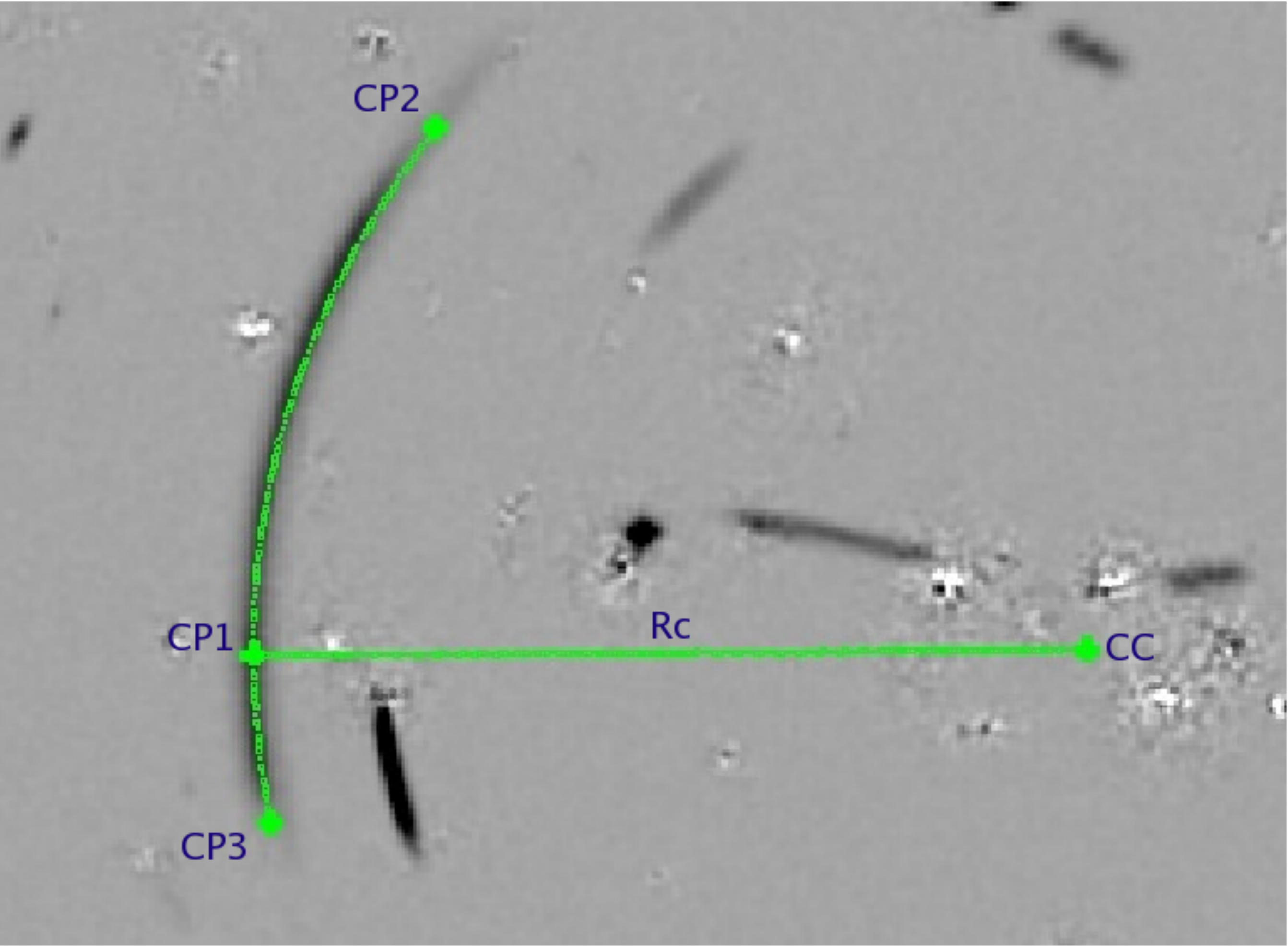} \includegraphics[width=0.52\hsize]{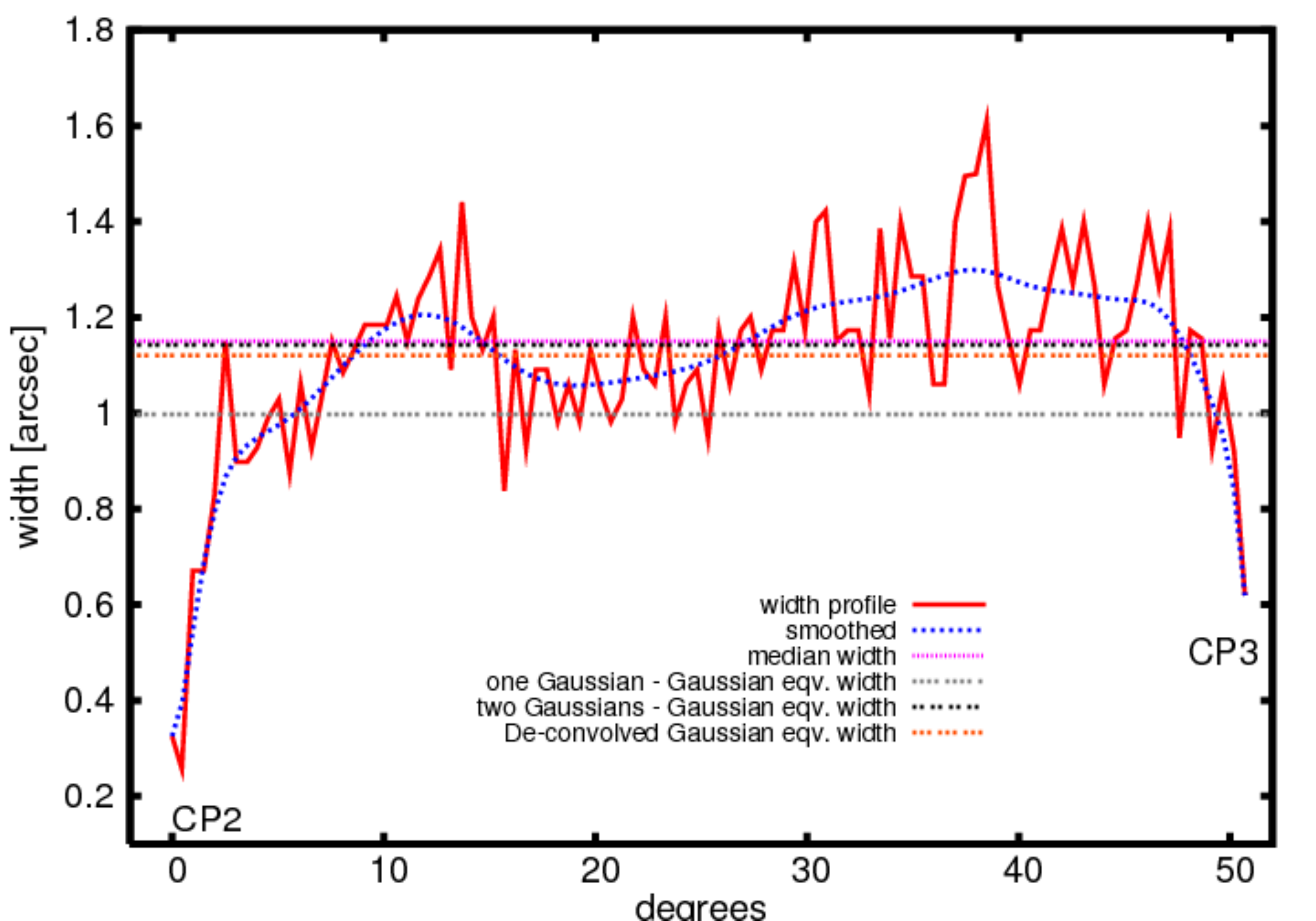}} 
\end{center}
\caption{Left panel: Fit of a gravitational arc. Three characteristic points have been identified on the image (CP1-3), through which a circular segment (CP2-CP3) has been traced. The curvature radius (Rc) and the center of curvature (CC) of the circular segment define the radius and the center of the arc. Right panel: Transversal width profile of the arc in the left panel (red solid line). The blue dotted line shows an interpolation of the profile with a Bezier curve. The horizontal lines correspond to various determinations of the arc width. Figures from \cite{2008A&A...482..403M}.}
\label{fig:simarc}
\end{figure}

While method 3 cannot be used on real data, as it requires to know the cluster radial and tangential magnification maps, even using method 1 and 2 result to be inaccurate, because of the noisy perimeter and area of the arc. \cite{2008A&A...482..403M} propose to measure the arc width by means of radial scans of the arc along straight lines passing through the center of curvature of the arc (CC, see Fig.~\ref{fig:simarc}) and intercepting the circular segment traced through the three characteristic points mentioned above (CP1, CP2, and CP3 in Fig.~\ref{fig:simarc}). At each scan, the maximal distance between arc points intercepted by the straight line is measured, thus constructing a transversal width profile of the arc. An example of such radial scansion is shown in Fig.~\ref{fig:simarc}. In the left panel, a simulated observation of a gravitational arc, obtained by strongly lensing a galaxy in the Hubble-Ultra-Deep field is displayed. The red line in the right panel indicates the transversal width profile of the arc along the circular segment CP2-CP3. The arc width is then defined as the median value of the width profile (magenta horizontal dotted line). \cite{2008A&A...482..403M} also discuss how to correct the arc width for PSF effects by Gaussian de-convolution (other horizontal lines).
\cite{2012A&A...547A..66R} recently used this method in ray-tracing simulations and compared it to the method 1 above. They found that, when applied to arcs originated from simple elliptical sources, the two methods lead to very similar results. 

\cite{2013A&A...549A..80F} recently presented some tools to model and fit gravitational arcs. Their {\tt ArcEllipse} method models arc-like images analytically expressing arc shapes as the distortion of ellipses such as one of their main axes is bent into an arc of a circle.    
Following this idea, the arc contour can be parametrized as
\begin{equation}
r_{\pm}(\theta)=r_c\pm \frac{W}{2}\sqrt{1-\left[2\frac{r_c(\theta_0-\theta)}{L}\right]^2} \;,
\label{eq:furl1}
\end{equation}
where $r_c$ is the curvature radius, $\theta_0$ is the position angle of the arc center, and $L$ and $W$ are the {\em equivalent} length and the width of the arc, respectively. These correspond to the major and minor axes of an ellipse having the same area of the arc. The arc is thus characterized by four parameters. To allow for asymmetries in the arc shape, \cite{2013A&A...549A..80F} suggest to split the length into the sum of a ``right-hand" and of a ``left-hand" lengths:
\begin{equation}
	L=b\left(\frac{1}{1-e_1}+\frac{1}{1-e_2}\right) \;.
\end{equation} 
This introduces an additional fifth parameter to characterize the arc shape. To define the arc position, one  need additional two coordinates, e.g. the center of curvature of the arc, $(x_0,y_0)$. 

Again following the prescription of \cite{2013A&A...549A..80F}, the brightness profile of the arc can be modeled assuming that the source generating the arc has an intrinsic \cite{1968adga.book.....S} brightness profile,
\begin{equation}
I_S(r)=I_e\exp{\left\{-b_n\left[\left(\frac{r}{r_e}\right)^{1/n}-1\right]\right\}} \;,
\end{equation}
where $I_e$ is the intensity at the effective radius $r_e$ that encloses half of the total light. Identifying $W$ with $2r_e$, the ``distorted" brightness distribution  of the arc is written as
\begin{equation}
	I(r,\theta)=I_0\exp{\left\{-b_n\left[\frac{\sqrt{[r_c(\theta-\theta_0)(1-e)]^2+(r-r_c)^2}}{r_e}\right]^{1/n}\right\}} \;.
	\label{eq:furl2}
\end{equation}
The normalization $I_0$ is given by $I_0=I_e e^{b_n}$. 

Aiming at quantifying the shape of an observed arc, one can then use the combination of Eqs.~\ref{eq:furl1} and \ref{eq:furl2} to perform a fit with nine free parameters ($x_0$,$y_0$,$r_c$,$\theta_0$,$e_1$,$e_2$,$r_e$,$n$,$I_0$). Note that  $b_n$ and $n$ are not independent\citep[see e.g.][]{1999A&A...352..447C}. 

\cite{2013A&A...549A..80F} apply this  technique to both real and simulated arcs. They quantify how well their {\tt ArcEllipse}+S\`ersic models can fit the data by means of the fit residuals. Not surprisingly, they find that the model reproduces very well the signal of the simulated arcs. However, it tends to over-estimate the signal of both arcs observed with HST or from the ground (e.g. arcs taken from the Canada-Frace-Hawaii-Telescope-Legacy-Survey). These results indicate that further improvements are needed in the method, which, nevertheless, represents a very promising approach. Testing it with more realistic simulations will help achieving better results \citep[see e.g. the recent work by][]{2012arXiv1212.1799B}.

The correlation between Einstein radii and lensing cross sections  discussed in Sect.~\ref{sect:einst} seem tight enough to infer sufficiently accurate estimates of optical depths for giant arcs from cosmological distributions of Einstein radii. This approach is computationally far less demanding than explicitly calculating individual lensing cross sections, since the computation of Einstein radii can be implemented in a particularly efficient way. Even from the observational point of view, measuring Einstein radii may be preferable to counting arcs. Indeed, parametric strong-lensing mass-modeling methods  were proven to be very efficient to constrain the matter distribution at the location of the critical lines \citep{2010A&A...514A..93M}. Therefore, studying Einstein Rings statistics, rather than arc statistics, should be seriously considered as a more efficient way to constrain cosmological parameters and structure formation with strong-lensing. 

\subsection{Modeling the lenses}
As discussed in Sect.~\ref{sect:complexsl}, several theoretical studies on arc-statistics helped to understand that strong-lensing clusters are particularly complex systems. This class of clusters cannot be described as a whole by means of simple analytical models like axially symmetric or even elliptical lenses. Indeed, it was found that several cluster properties contribute to the strong-lensing cross sections: cluster galaxies (in particular the BCGs), substructures, asymmetries and triaxiality of the mass distributions, mergers, effects of baryons, etc. This given, the only possible way-out for reliable arc statistics studies appeared to model cluster lenses numerically, via N-body/hydrodynamical simulations. Aiming at investigating the dependence of arc-statistics on cosmological parameters, this is however a strong limitation. The advent of increasingly faster supercomputers helps, of course, but the realization of simulations exploring continuous ranges of parameter values and  their subsequent analysis via ray-tracing methods is overwhelmingly demanding. Additionally, because of the fact that massive clusters are the rarest bound structures in the universe, large cosmological volumes need to be sampled in order to limit the impact of cosmic variance.

\begin{figure}[t]
\begin{center}
\resizebox{12cm}{!}{\includegraphics[width=0.4\hsize]{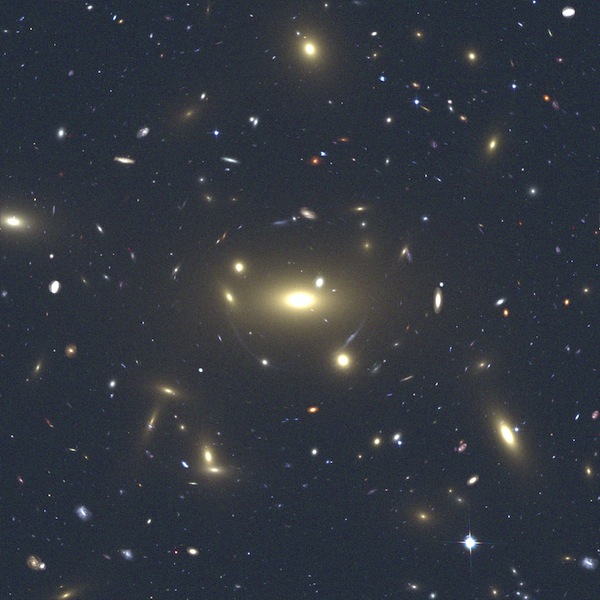} \includegraphics[width=0.59\hsize]{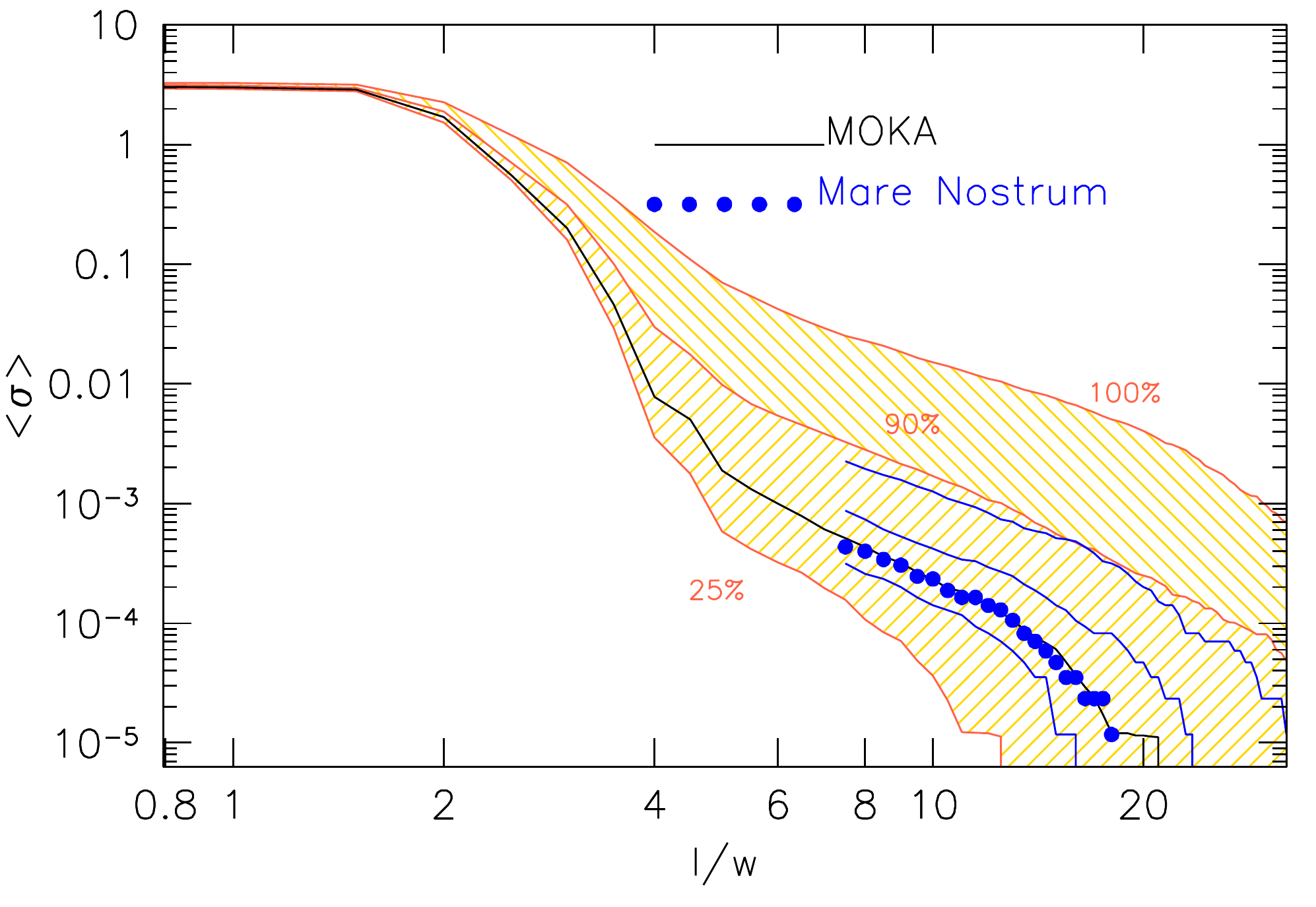}} 
\end{center}
\caption{Left panel: simulated observation with HST of a mock galaxy cluster produced with the {\tt MOKA} code \citep{2012A&A...547A..66R}. The simulated observation is obtained by combining three images in the F475W, F606W, and F814W filters of HST, using the code {\tt SkyLens} \citep{2008A&A...482..403M}. Right panel: comparison  between the median lensing  cross section
  $\sigma$ of halos modeled with {\tt MOKA} (solid line) and of halos extracted from the {\sc MareNostrum Universe} (filled circles). Results are shown as function of the minimal length-to-width ration $l/w$ and for a sample of cluster-sized haloes in
  the mass range $6-7  \times 10^{14} h^{-1} M_{\odot}$.  The shaded regions
  enclose $25\%-90\%$  and $90\%-100\%$ for a  sample of \texttt{MOKA}
  haloes, while the three solid curves illustrate the same regions for
  the haloes in the {\sc MareNostrum Universe}. Figure from \cite{2012A&A...547A..66R}.}
\label{fig:moka}
\end{figure}

For these reasons, few authors are pursuing a compromise between fully numerical and fully analytical methods. They use the results of large cosmological simulations to construct realistic mass distributions via combinations of analytical mass components. An example is given the public code {\tt MOKA} \citep{2012MNRAS.421.3343G}. This code construct realizations of cluster lenses combining 1) a smooth halo component, resembling the dark matter halo of the cluster; 2) a central concentration of ``stars", resembling the presence of a massive galaxy dominating the mass distribution in the cluster center; and 3) a number of substructures, which constitute the clumpy component of the cluster mass distribution. The smooth halo has a triaxial shape, with triaxiality assigned using the recipe of \cite{JI02.1}, which is calibrated on numerical simulations. The density profile of the smooth halo is modeled using a generalized NFW profile (Eq.~\ref{eq:gnfw}).  
The halo concentration, $c=r_{vir}/r_{s}$, is chosen such to mimic the $c-M$ relation proposed by \cite{2009ApJ...707..354Z}, and scatter is added assuming that concentrations at fixed mass are log normally distributed. Stars in the BCG are distributed according to an \cite{1990ApJ...356..359H} or to a \cite{JA83.1} density profile. The code implements also the recipe of \cite{1986ApJ...301...27B} to mimic the adiabatic contraction of the smooth dark matter halo, due to the growth of the central galaxy. The substructures are distributed reproducing the mass function and the radial distribution found by \cite{2004MNRAS.355..819G} studying dark matter halos evolved in large cosmological volumes. Galaxy properties like the dark matter and the stellar masses, luminosities, etc,  are determined using the halo occupation distribution approach \citep{2006MNRAS.371..537W}. The substructure density profiles are modeled using truncated isothermal spheres \citep{1998MNRAS.299..728T}. An example of cluster realization obtained with {\tt MOKA}, of which HST observations have been simulated with the {\tt SkyLens} software \citep{2008A&A...482..403M}, is shown in the left panel of Fig.~\ref{fig:moka}. 

Testing the code against halos in the {\sc MareNostrum Universe}, \cite{2012MNRAS.421.3343G} found that halos simulated with {\tt MOKA} reproduce very well the lensing cross sections of fully numerical halos (see right panel in Fig.~\ref{fig:moka}). This is a very reassuring  result, which shows that lens models constructed with this semi-analytic approach are sufficiently realistic.

Also \cite{2012A&A...547A..66R}  used semi-analytic recipes to build mock lenses. Although, their mass models are less sophisticated than those produced by {\tt MOKA}, they implemented an interesting method, based on theoretical merger-trees, to include  merger dynamics in simulated light cones. Very shortly, their method consists of generating a population of halos at redshift zero, which is then evolved back in time using merger trees, following the approach proposed by \cite{2008MNRAS.389.1521Z}. In this way, a single halo at redshift zero can be split into its smaller progenitors at any given observing time. The merger kinematics are simulated estimating the duration of the event as given by the dynamical time scale
\begin{equation}
T_{\rm dyn} = \sqrt{\frac{(r_{\rm vir,1}+r_{\rm vir,1})^3}{G(M_1+M_2)}} \;,
\end{equation}  
where $M_i$ and $r_{\rm vir,i}$ are the virial masses and radii of the merging components. The direction of the motion is taken to be random and the velocity is calculated  assuming a uniform linear motion. As a first application of this algorithm, \cite{2012A&A...547A..66R} compared the distributions of the largest Einstein radii including and excluding cluster mergers, finding that mergers increase the theoretically expected number of Einstein radii above $10"$ and $20"$ by 36\% and 74\% respectively. 
 
\subsection{Observations}
In the next years, an increasing number of observational campaigns will explore nearly the full sky looking for galaxy clusters at all wavelengths where they can be detected. For example, the eRosita satellite is expected to compile a catalog of 50-100 thousand X-ray clusters by  performing a deep survey of the entire X-ray sky \citep{2012arXiv1209.3114M}. These data will be a complement to the SZ observations delivered by the Planck mission and by other microwave/millimeter surveys like those operated with the ACT and SPT. Additionally, several optical surveys are starting now or will start soon (KIDS\footnote{\tt http://kids.strw.leidenuniv.nl}, DES\footnote{\tt http://www.darkenergysurvey.org/}), which has been designed for the (weak-)lensing analysis of thousands of square degrees of the sky. On a ten-years time-scale, the Large-Synoptic-Survey-Telescope (LSST\footnote{\tt http://www.lsst.org/}) and the Euclid satellite\footnote{\tt http://www.euclid-ec.org/} will also start their operations. New frontiers will be open for statistical applications of  strong lensing with samples of clusters selected in different ways.

Despite they are not fully designed for this purpose, all the above mentioned optical surveys will enable strong-lensing  science on the cluster scales.  In particular, among them, the Euclid space mission will provide  the highest quality data  and the widest sky coverage. The main survey of Euclid will cover an area of 15,000 square degrees of  the sky  at galactic  latitudes $|b|>30\deg$. In the visual band, the telescope will be sensitive to photons at wavelengths between  $0.55$ and $0.9 \mu$m. For the same fields, Euclid will  deliver imaging in the $Y,J,H$  NIR bands as well as slitless spectra covering  the wavelength range $1.1-2.0 \mu$m. Complementary data will be available through the synergy with several ground based facilities \citep{2011arXiv1110.3193L}. Given the  good resolution (0.101"/pixel)  and sensitivity ($24.5$ AB magnitudes in the visual band),  Euclid will  be able  to  detect and resolve  with  sufficient  accuracy  a huge number of strong  lensing  features, such as gravitational arcs,  arising from highly magnified distant galaxies near the cluster cores, and use them in combination with weak-lensing to extend the mapping of the cluster content to the central regions. This will enable to perform arc and Einstein ring statistics studies with unprecedented accuracy.

\begin{figure}[t]
\begin{center}
\resizebox{12cm}{!}{\includegraphics[width=0.49\hsize]{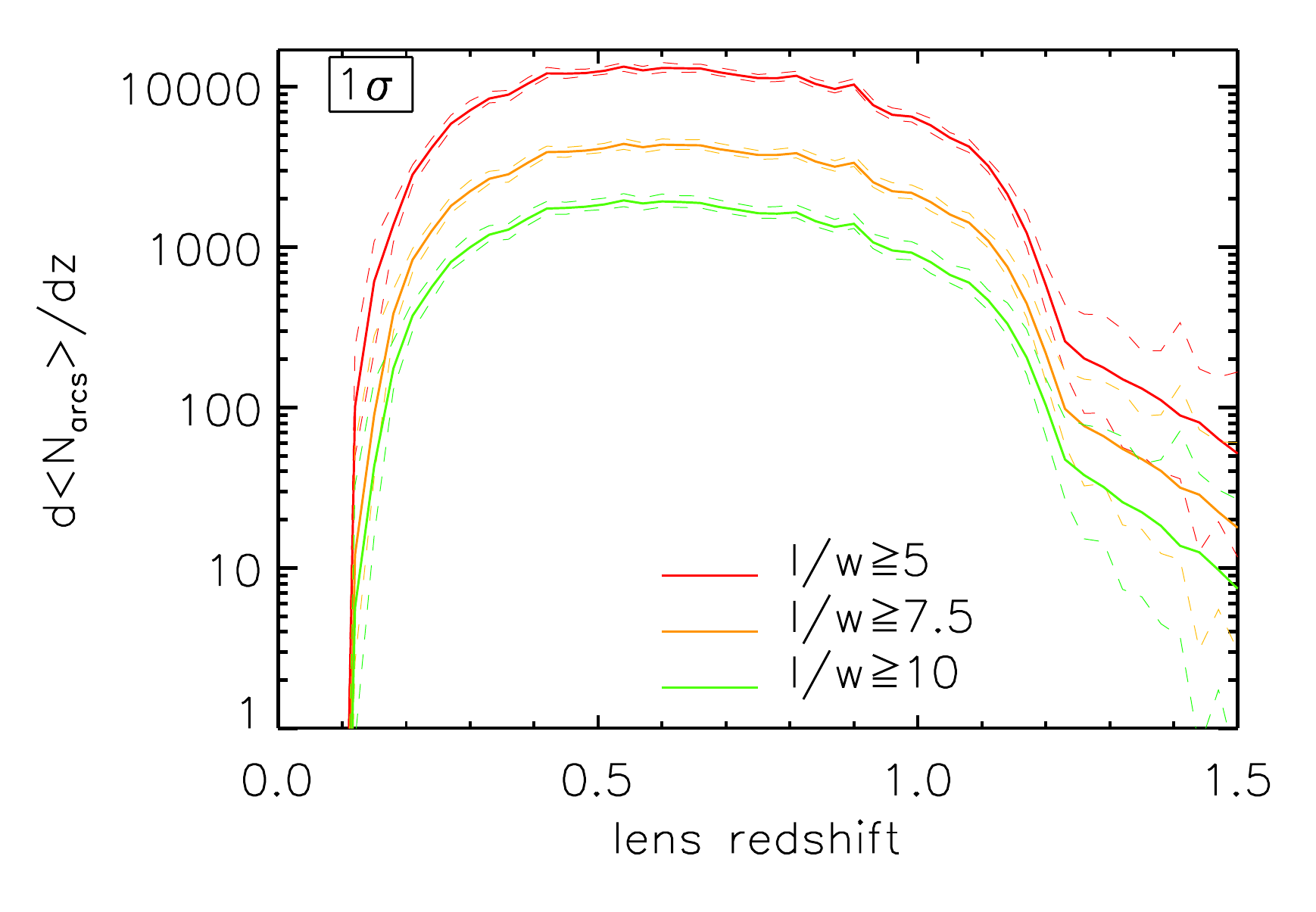} \includegraphics[width=0.49\hsize]{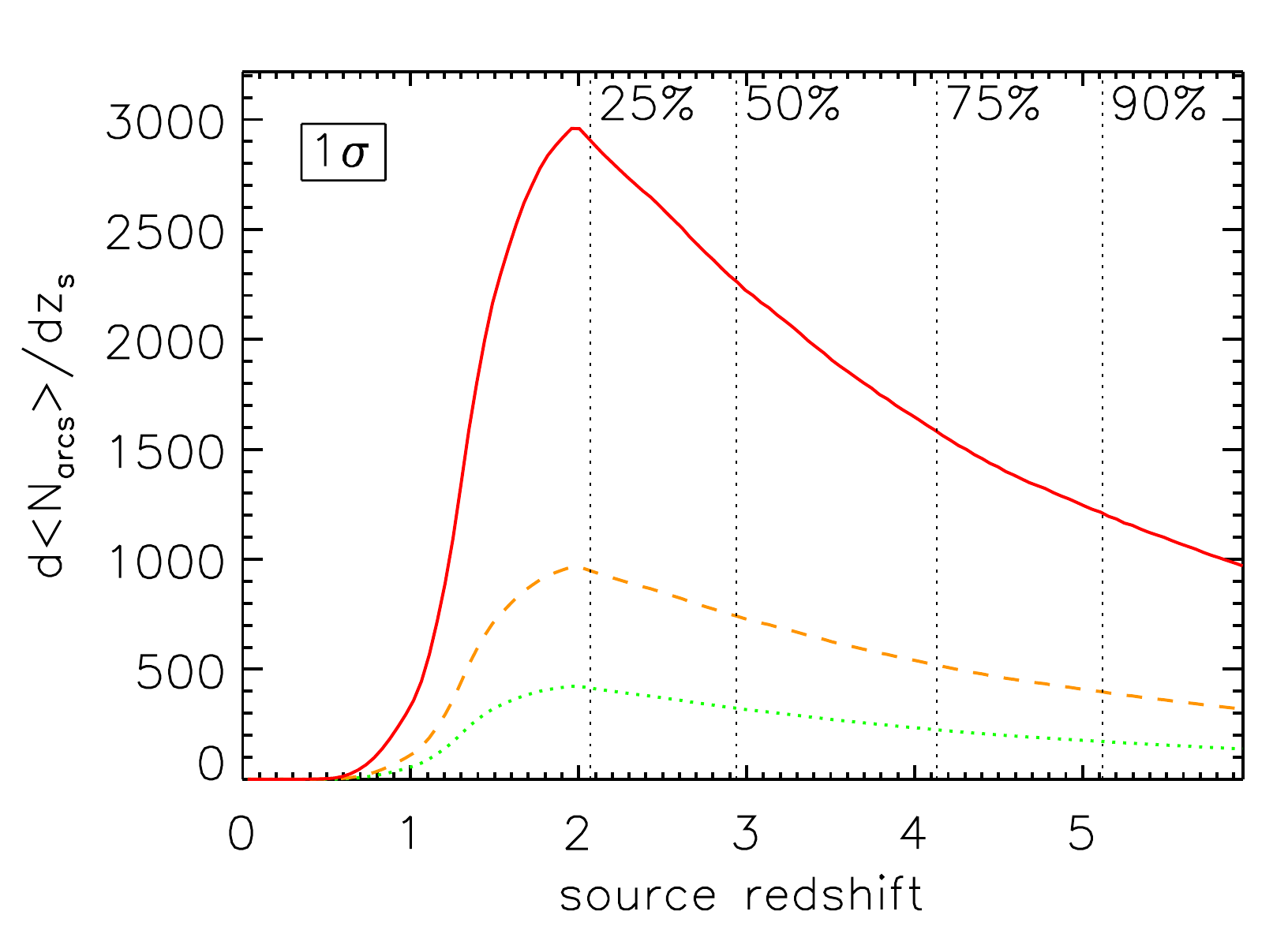}} 
\end{center}
\caption{Left panel: number of arcs expected to be detectable in the Euclid wide-survey as a function of the lens redshift. The thick
   solid red,  orange, and green lines  are the medians  among the 128
   survey realizations  and they refer  to arcs with  $l/w\geq 5$,
   $7.5$, and $10$, respectively. Right panel: Number of  arcs as  a function of  source redshift.  Again, results are  shown for arcs with $l/w\geq 5$
   (solid  line), $7.5$ (dashed  line), and  $10$ (dotted  line).  The
   fractional numbers of arcs originated from sources at $z\leq z_{\rm
     s}$ are given by the  vertical dotted lines: they are independent
   of the value of $l/w$. Figure from \cite{2012MNRAS.427.3134B}.}
\label{fig:boldrin}
\end{figure}

The feasibility of arc statistics studies with Euclid has been recently investigated by \cite{2012MNRAS.427.3134B}. They used the {\tt MOKA} code to generate past-light cones mimicking the survey which will be performed by Euclid, assuming a $\Lambda$CDM cosmology, and they used Euclid image simulations to quantify the redshift evolution of the sources which will be detectable in typical Euclid images. Using ray-tracing techniques, they distorted these galaxies according to the {\tt MOKA} deflection angle maps and estimated that the number of arcs which will emerge above the sky background in the Euclid wide-field survey will be $8912_{-73}^{+79}$,  $2914_{-25}^{+38}$, and  $1275_{-15}^{+22}$ for $l/w\ge5$, $7.5$  and $10$, respectively. They also found that most of  the lenses which will contribute to the  lensing optical depth are located at redshifts $0.4<z<0.7$ and  that the $50\%$ of the  arcs are images of  sources at  $z>3$ (see Fig~\ref{fig:boldrin}).
This experiment shows that 1) it is possible to simulate  full sky surveys in a reasonable time (a week time-scale) with the current  tools for lens modeling and  the existing super-computing machines. This suggests that it is becoming feasible to study the evolution of arc statistics as a function of several cosmological parameters; and 2) Euclid  have the potential to increase the number of the known giant arcs by a factor of $\sim 30$. This huge size increment of the observational data-set will likely  limit significantly the uncertainties
 of the arc-statistics approach.

A potentially very interesting, but still relatively unexplored field where arc statistics could find interesting applications is that of the submm galaxies. This topic is discussed in a paper by \cite{2009A&A...508..141F}. As pointed out by \cite{1996MNRAS.283.1340B,1997MNRAS.290..553B}, the fraction of lensed sources observed in the mm/submm wavebands is expected to be much larger than in the optical. Indeed, due to the spectral shape of the thermal dust emission, the observed submm flux density of dusty galaxies with a given luminosity remains approximately constant in the redshift range $1 \lesssim z \lesssim 8$ instead of declining with increasing distance. This effect, together with the steep slope of the observed submm number counts, produces a strong magnification bias that makes submm galaxies  an ideal source population for the production of lensed arcs. \cite{2009A&A...508..141F} estimated that  a submm all-sky survey with a sensitivity of 1 mJy arcsec$^{-2}$ will detect hundreds of arcs with a 5$\sigma$ significance. In the radio, this number can be achieved with a sensitivity of 10-20 $\mu$Jy arcsec$^{-2}$. The advent of ALMA will open a new window into mm/submm astronomy at sub-arcsecond resolution and sub-mJy sensitivity, allowing the detection of resolved gravitational arcs produced by submm galaxies. However a survey capable of producing a sufficient number of detections for arc statistics, i.e. being a good compromise between area, depth, and resolution
 still has to be designed.
  
\subsection{Arc-finders}
\label{sect:arcfinders}
Current and future surveys will cover large fractions of the sky delivering huge amounts of data where strong lensing features will be hidden. It is not feasible to systematically look for these features by means of human inspection. For this reason, several authors began to develop tools for the automatic search of gravitational arcs. 

Since the beginning it appeared that properly identifying gravitational arcs, while keeping the level of contamination at reasonable levels  is a difficult task. One algorithm has been developed by \cite{2004A&A...416..391L}. It uses anisotropic diffusion tensors to identify the extended and highly anisotropic arcs. This feature causes it to create arcs from random noise, which is of course unwanted.

An alternative has been developed in the CFHTLS team \citep{AL06.1}. The method consists of measuring a local value of the elongation and of the orientation of the surface brightness distribution at each point on the astronomical images, convolved with a smoothing kernel designed to enhance the detection of arcs. Adopting a threshold, elongated structures are identified and reconstructed using connectivity criteria. This algorithm has been used in the SL2S survey \citep{CA07.1,2012ApJ...749...38M}. When applied to the data, the automatic detection produced $\sim 1000$ candidate arcs over an area of $\sim 150$ sq. degs. A first visual inspection reduced the detection to $413$ candidates, which were subsequently considered for ranking by three people. The candidates that were finally classified as arcs were 127 (12 giant arcs with $l/w>8$). They were found at distances $\sim 2''-18''$ from the centers of their lenses (mostly galaxy groups) and to have a mean (photometric) redshift $z\sim 0.6$. The false detections during the automatic search were prevalently due to spikes and halos near stars. 

\begin{figure}[t]
\begin{center}
\resizebox{12cm}{!}{
\includegraphics[width=0.49\hsize]{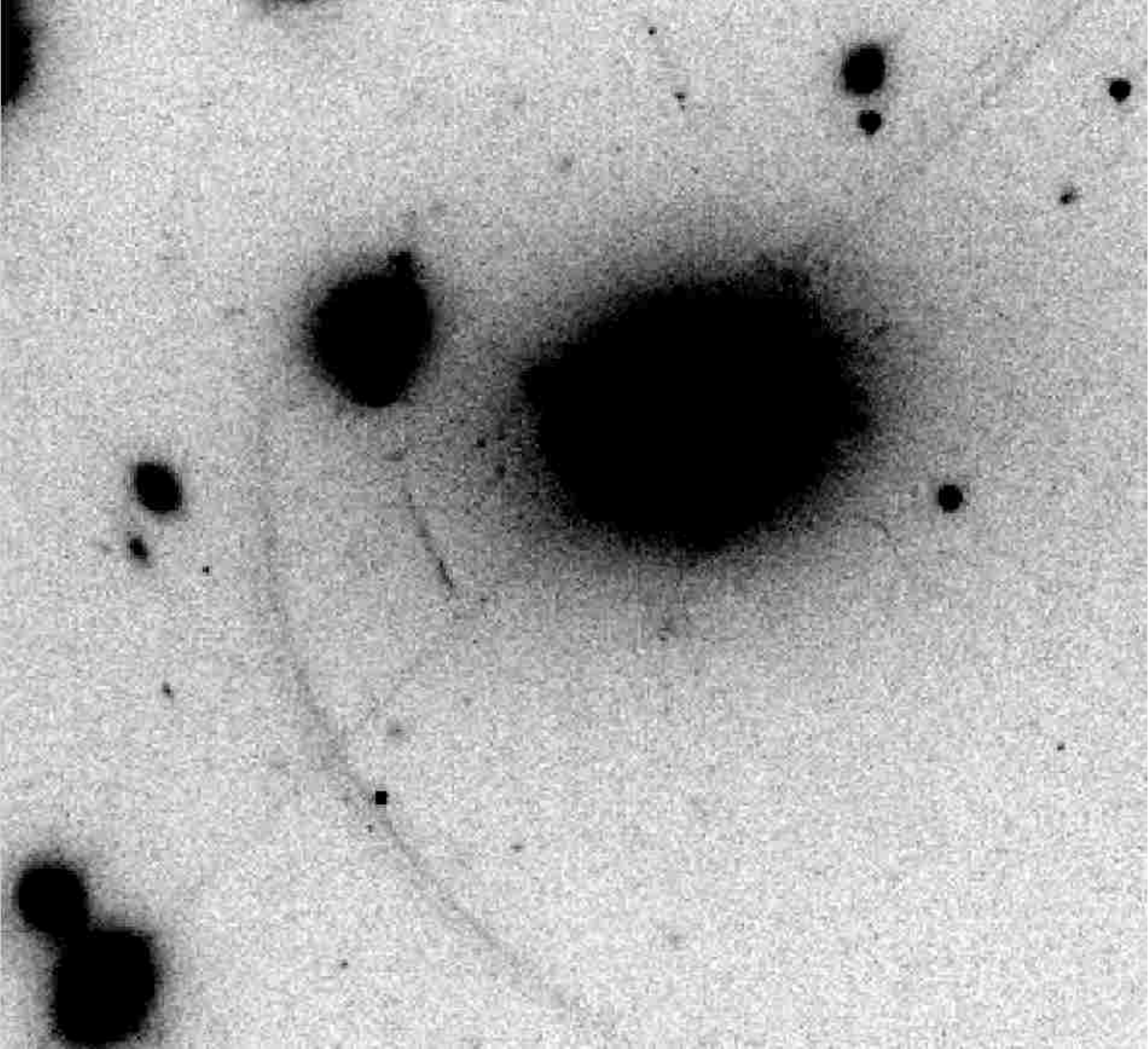} 
\includegraphics[width=0.49\hsize]{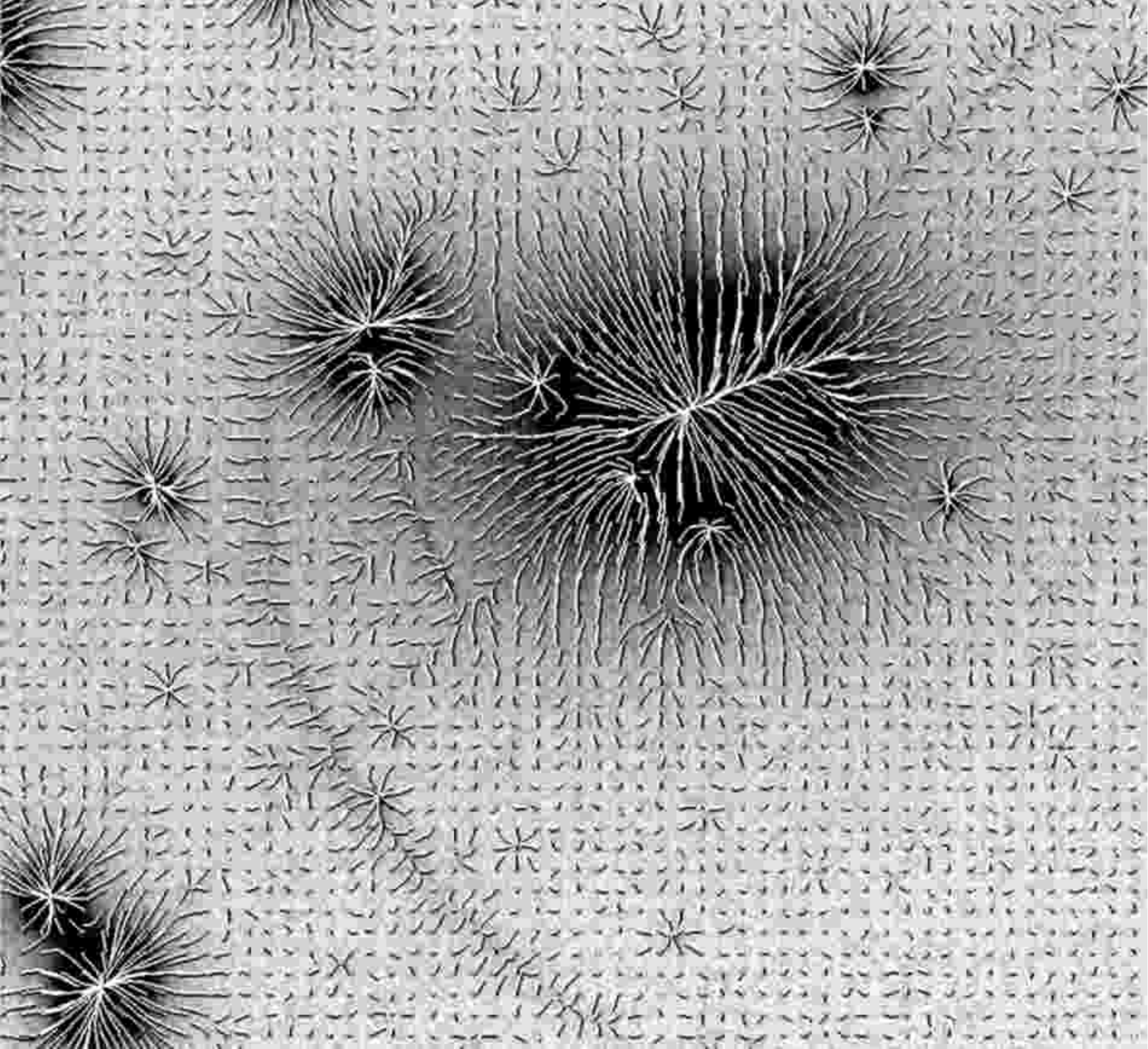}
\includegraphics[width=0.49\hsize]{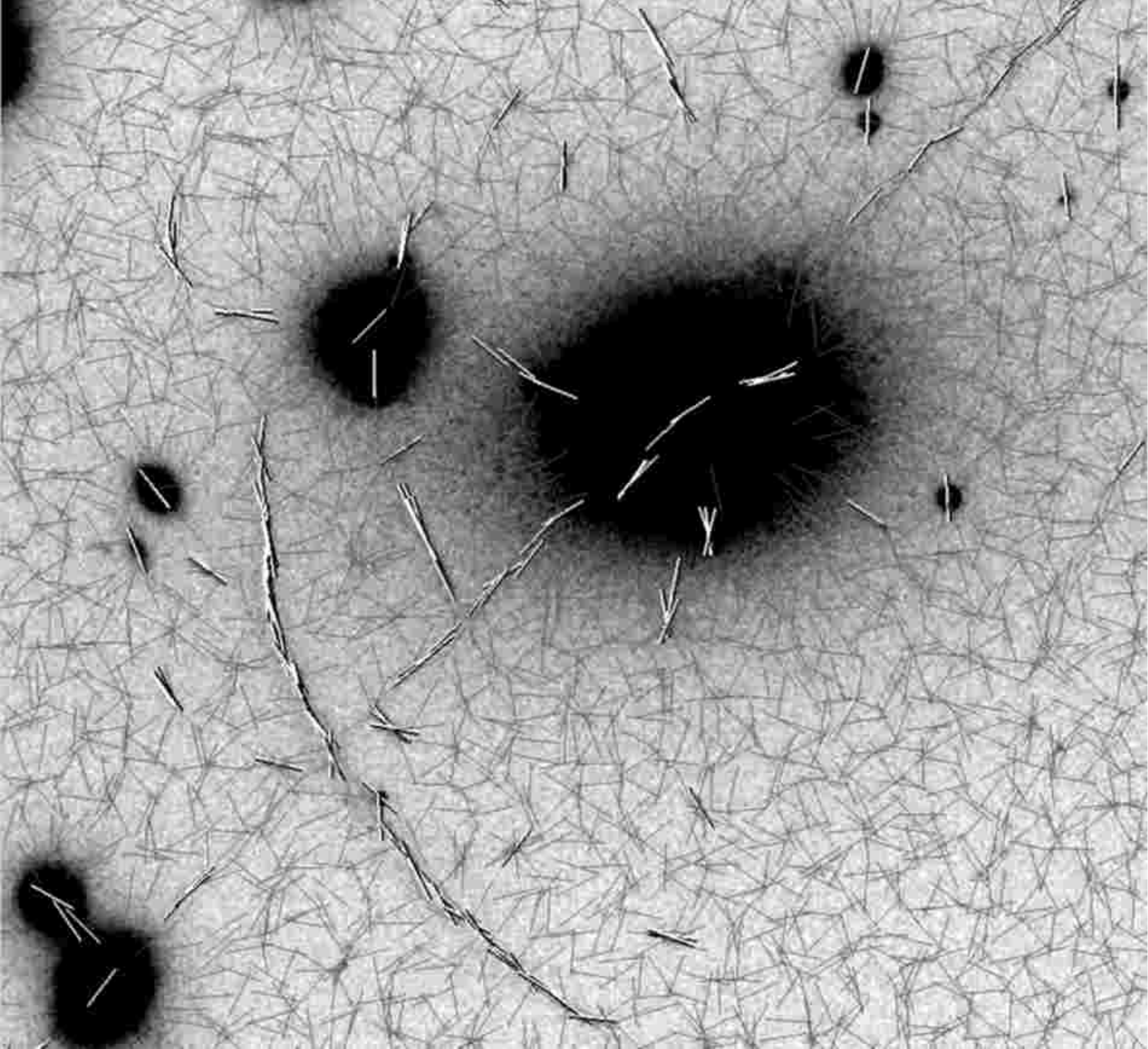}
\includegraphics[width=0.49\hsize]{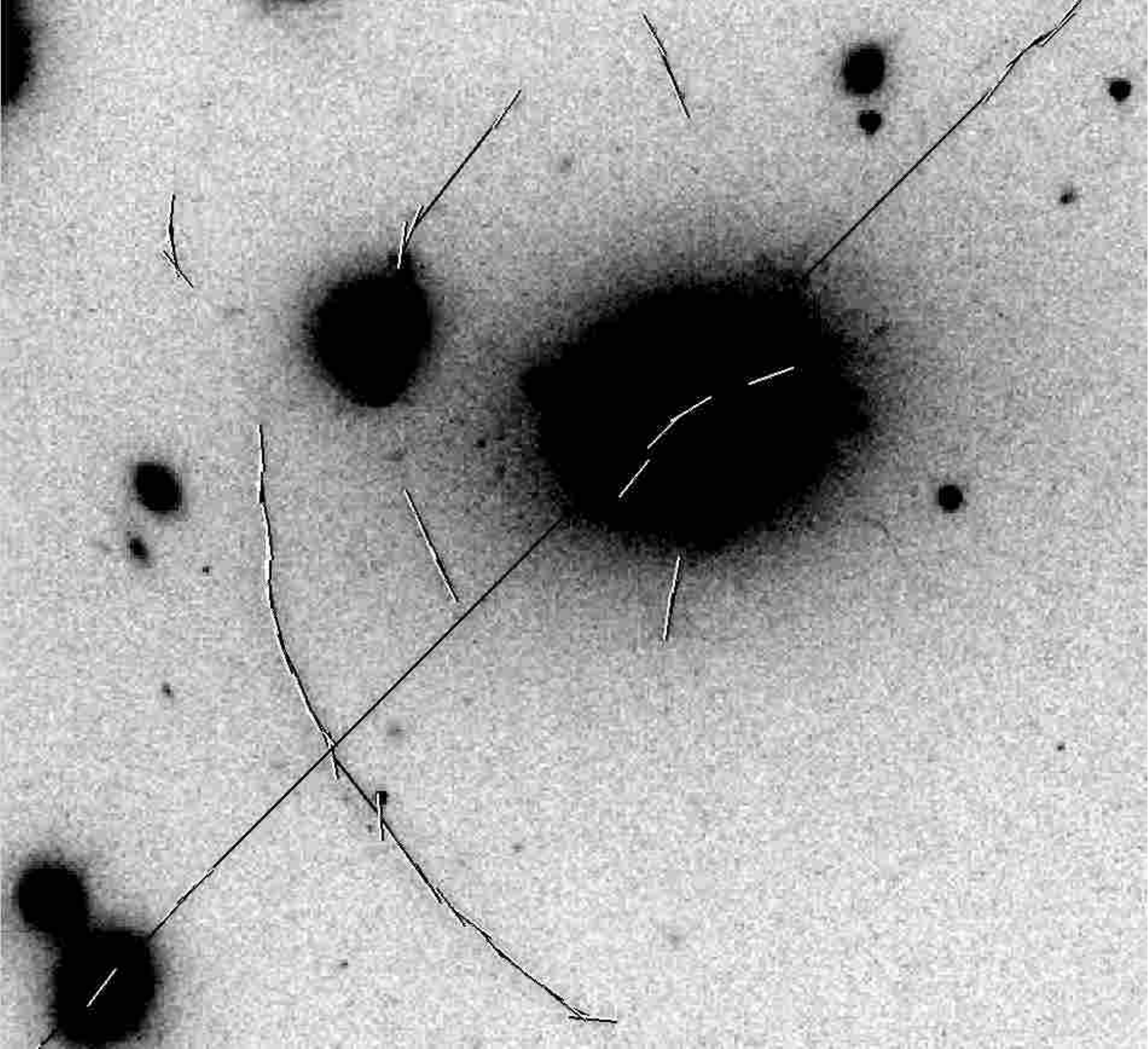}
} 
\end{center}
\caption{The procedure of \cite{SE07.1} to detect arcs applied to the core of the galaxy cluster A2390. From left to right: 1) the astronomical image; 2) the  subsections, originally covering the image uniformly, are moved, leading to the closest local centers of brightness; 3) subsections are oriented according to the local brightness distribution; 4) arcs are identified as collections of subsections with correlated orientations.}
\label{fig:arcfinder}
\end{figure}

Other algorithms have been presented  by \cite{HO05.1} and by \cite{SE07.1}. One is based on a combination of the two software packages SExtractor and IRAF and identifies arcs as faint extended sources until just above the noise limit. The other is specifically developed to detect extended images near the noise threshold without any convolution applied. It is based on the  idea that arcs are coherent structures extending over many pixels with a local brightness maximum. The procedure is described in Fig.~\ref{fig:arcfinder} and  consists of i) covering the astronomical image with equally distributed small subsections; ii) move the subsections iteratively towards local centers of intensity; iii) compute the ellipticity and the orientation of each subsection; and iv) detect arcs where the subsection orientations are correlated, taking into account their relative position. The algorithm is capable of searching through large images quickly and thus to process wide-field data in a short time. Besides, its application is followed by a post-processing step removing a good fraction of the spurious detections from the final arc list. Nevertheless, spurious detections cannot be fully suppressed.
So far, the level of completeness and contamination reached by all these techniques were not quantified by means of realistic simulations, but their application to the data show that these algorithms are potentially very efficient to find gravitational arcs. 

\section{Conclusions}
In this review, we discussed several important results that emerged in a number of studies since the arc statistics was proposed as a tool for cosmology. From the beginning, the arc statistics approach produced puzzling results and stimulated a strong debate on whether an inconsistency between the expected number of {\em giant} gravitational arcs in the so called concordance model is compatible with observations or not. After more than fifteen years, a solution to the arc statistics problem seem still far away. The debate on this issue has been reinvigorated recently by other observations of strong-lensing clusters, which show that these are way too concentrated and produce way too large Einstein radii compared to theoretical expectations. 

From the existing literature, several interesting conclusions can be drawn:
\begin{itemize}
\item arc statistics is still characterized by too large uncertainties for being a reliable and robust tool for cosmology. However, the attempts to understand if the arc statistics problem could be solved by means of previously neglected and important cluster properties have significantly enhanced our comprehension of the internal structure of strong lensing galaxy clusters. They taught us that a correct modeling of the cluster lenses must take into account the effects of ellipticity, asymmetries, substructures, cluster galaxies, and, mostly important, dynamical events like mergers between mass sub-components. Thus, arc statistics has significantly contributed to our understanding of the strong lensing cluster population;  
\item the strong sensitivity of the strong lensing cross sections to merger events, enables to use the arc statistics as a tracer of the evolution of the cosmic structures. The rate of production of giant arcs as a function of redshift should reflect the rate of mergers between substructures as a function of the cosmic time;
\item one of the largest difficulty in the application of arc statistics in cosmology is the lack of a proper interface between theory and observations. A lot of work has been done in the last years in order to improve this situation, but in the meantime it became clearer that there might be alternatives to simply counting arcs in order to exploit the sensitivity of the strong lensing signal to cosmology. For example, the cross section for giant arcs is strongly correlated to the size of the lens Einstein radii. This seems to be measurable in observations with a good precision, given the existing parametric mass modeling techniques. Additionally Einstein radii can be computed quickly and efficiently also in numerical simulations;
\item a big limitation in the application of arc statistics for cosmology was represented by the very limited number of observed gravitational arcs to be compared to theoretical predictions. In the last years, several surveys began which are focussed on robustly estimating the frequency of giant arcs behind galaxy clusters. To date, the number of known giant arcs amounts is of the order of $\sim 100$ arcs. A big boost of the size of the observational data-sets is expected in the next years thanks to the upcoming (nearly) all sky surveys at several wavelengths. These will enable to 1) to increase the statistical significance of the comparison between theory and observations, even thanks to the development of specific tools for the automatic detection of gravitational arcs in wide field images, and 2) to better identify the selection functions to be applied to both observed and simulated data;
\item our good understanding of what are the cluster properties that are affecting the cluster cross sections for strong lensing, combined with the huge advancements in computational power, allow us to construct sophisticated and reliable tools for calculating the expected strong lensing signal as a function of the cosmological parameters. Simulating all sky surveys in a given cosmological framework is now possible within relatively short time-scales using the above mentioned tools.      
\end{itemize}  
All of this, further supports our sentiment that the arc statistics and its spin-offs, like the Einstein ring statistics, will be a fundamental tool for cosmology in the very near future.

\begin{acknowledgements}
We are grateful to the International Space Science Institute of Bern for supporting us in the work for the preparation of this manuscript. M.M. acknowledges support from PRIN INAF 2009 and ASI (agreement Euclid phase B2/C).  
\end{acknowledgements}

\bibliographystyle{aps-nameyear}      % basic style, author-year citations
\bibliography{../../TexMacro/master}   % name your BibTeX data base
\nocite{*}

\end{document}